\documentclass[12pt]{article}
\usepackage{graphics,epsfig}

\def\dalemb#1#2{{\vbox{\hrule height .#2pt
        \hbox{\vrule width.#2pt height#1pt \kern#1pt
                \vrule width.#2pt}
        \hrule height.#2pt}}}

\let\a=\alpha \let\b=\beta \let\g=\gamma \let\d=\delta \let\e=\epsilon
\let\z=\zeta  \let\th=\theta  \let\k=\kappa
\let\l=\lambda \let\m=\mu  \let\x=\xi \let\p=\pi 
\let\s=\sigma \let\t=\tau   \let\c=\chi 
 \let\vep=\varepsilon
\let\w=\omega       \let\D=\Delta \let\Th=\Theta \let\L=\Lambda
\let\X=\Xi \let\P=\Pi \let\S=\Sigma  \let\Y=\Psi
\let\C=\Chi \let\W=\Omega
\let\la=\label \let\ci=\cite 
  
\def\nn{\nonumber} \def\bd{\begin{document}} \def\ed{\end{document}}
\def\ds{\documentstyle} \let\fr=\frac \let\bl=\bigl \let\br=\bigr
\let\Br=\Bigr \let\Bl=\Bigl
\let\bm=\bibitem
\let\na=\nabla
\def\tU{{\widetilde U}}
\let\pa=\partial \let\ov=\overline
\def\ie{{\it i.e.\ }}
\newcommand{\be}{\begin{equation}}
\newcommand{\ee}{\end{equation}}
\def\ba{\begin{array}}
\def\ea{\end{array}}
\def\ft#1#2{{\textstyle{{\scriptstyle #1}\over {\scriptstyle #2}}}}
\def\fft#1#2{{#1 \over #2}}
\def\F#1#2{{ F_{#1}^{(#2)} }}
\def\cF#1#2{{ {\cal F}_{#1}^{(#2)} }}

\def\={\, =\, }
\def\+{\, +\, }
\def\-{\, -\, }

\def\R{{\bf R}}
\def\sst#1{{\scriptscriptstyle #1}}
\def\oneone{\rlap 1\mkern4mu{\rm l}}
\def\e7{E_{7(+7)}}
\def\td{\tilde}
\def\wtd{\widetilde}
\def\im{{\rm i}}
\newcommand{\ho}[1]{$\, ^{#1}$}
\newcommand{\hoch}[1]{$\, ^{#1}$}
\newcommand{\bea}{\begin{eqnarray}}
\newcommand{\eea}{\end{eqnarray}}
\newcommand{\ra}{\rightarrow}
\newcommand{\lra}{\longrightarrow}
\newcommand{\Lra}{\Leftrightarrow}
\newcommand{\ap}{\alpha^\prime}
\newcommand{\bp}{\tilde \beta^\prime}
\newcommand{\cB}{{\cal B}}
\newcommand{\cO}{{\cal O}}
\newcommand{\vecx}{\vec{x}}
\newcommand{\vecy}{\vec{y}}
\newcommand{\vecp}{\vec{p}}
\newcommand{\vecq}{\vec{q}}
\newcommand{\tr}{{\rm tr} }
\newcommand{\Tr}{{\rm Tr} }

\newcommand{\cL}{{\cal L}}
\newcommand{\cA}{{\cal A}}
\newcommand{\cD}{{\cal D}}
\def\sst#1{{\scriptscriptstyle #1}}
\def\0{{\sst{(0)}}}
\def\1{{\sst{(1)}}}
\def\2{{\sst{(2)}}}
\def\3{{\sst{(3)}}}
\def\4{{\sst{(4)}}}
\def\5{{\sst{(5)}}}
\def\6{{\sst{(6)}}}
\def\7{{\sst{(7)}}}
\def\8{{\sst{(8)}}}
\def\ve{\varepsilon}
\def\vf{\varphi}
\def\F{\Phi}
\def\wg{\wedge}

\def \nn {\nonumber}
\def \rk  {m}
\def \L {{\Lambda}}
\def \ka  { {\kappa }}
\def \S {{ \call S}}

\def \foot {\footnote}
\def \bi{\bibitem}

\def \tr {{\rm tr}}
\def \ha {{1 \over 2}}
\def \td {\tilde}
\def \ci{\cite}
\def \N {{\mathcal N}}
\def \ww {\Omega}
\def \const {{\rm const}}
\def \ss {\sum_{i=1}^3 }
\def \t {\tau}
\def\S{{\mathcal S} }
\def \XX {{\rm X}}

\def \lra {\leftrightarrow}
\def \vom {{\bar \omega}}
\def \E {{\mathcal  E}} \def \J {{\mathcal  J}}
\def \YY {{\rm Y}}

\def \d {\del}
\def \rJ {{J}}
\def \sms {sigma models\ }
\def \sm {sigma model\ }
\def \L {\Lambda}
\def \gl {\ell}
\def \tr {{\rm tr\ }}
\def\z{\zeta}
\def\zi{\zeta_1}
\def\zii{\zeta_2}
\def\K{\mbox{K}}
\def\eE{\mbox{E}}   \def \vt {\vartheta}
\def \vr {\varrho}
\def \wup {w}

\def\dg{\dagger}
\def\a{\alpha}
\def\b{\beta}
\def\e{\varepsilon}
\def\p{\phi}
\def\ap{\alpha^\prime}
\def\I{{\cal I}}

\def\R{{\bf R}}
\def\Z{{\bf Z}}
\def\C{{\bf C}}
\def\P{{\bf P}}
\def\xb{{\bar X}}
\def\Tr{{\rm  Tr}}
\def\tr{{\rm  tr}}

\def \del{\partial}
\def \a {\alpha}
\def \aa {{\a'}}
\def\g{\gamma}
\def\s{\sigma}
\def\z{\zeta}
\def\zi{\zeta_1}
\def\zii{\zeta_2}
\def\ov{\over}

\def\I{{\cal I}}
\def\J{{\mathcal J}}
\def \ok {{1\ov \k}}
\def\LL{{\mathcal L }}
\def \jL {{J}}
\def \om {\omega}
\def \cL {{\mathcal L}} \def \cH {{\mathcal H}}
\def\E{{\mathcal E}}
\def\w{\omega}
\def\b{\beta}
\def\l{\lambda}
\def\eps{\epsilon}
\def\vep{\varepsilon}
\def \De {{\mathcal D}}

\def  \Jt {  {J}_{\rm tot}    }

\def \k {\kappa}
\def\foot{\footnote}
\def \four{{\textstyle {1\ov 4}}}
 \def \third { \textstyle {1\ov 3
}}
\def\det{\hbox{det}}
\def \ci {\cite}

\def \foot {\footnote}
\def \bi{\bibitem}

\def \tr {{\rm tr}}
\def \ha {{1 \over 2}}
\def \tid {\tilde}
\def \vv {{\rm v}}
\def \tl {{\tilde \l}}
\def \XX {{\rm X}}
\def \ta {{\tilde \a}}
\def \fo { {1\ov 4}}
\def \ep {\epsilon}
\def \inti {{\int^{2\pi}_0 {d \sigma \ov 2 \pi}}}

\def \d {\partial}
\def \K {{\rm S}}
\def \el {\ell}
\def \Tr {{\rm Tr}}
\def \P {\Phi}
\def \l  {\lambda}
\def \tl {{\tilde \l}}
\def \bl {{\tilde \l}}
\def \const {{\rm const}}
\def \V {v}

\def \bv {v^*}
\def \vv {{\rm v}}
\def \LL {{\mathcal L}}
\newcommand{\PV}[1]{P_{\!\!_{V_{#1}}}}

\def \bL {\ell}
\def \M {{\mathcal M}}
\def \N {{\mathcal N}}
\def \S {{\rm S}}
\def \vn {\vec n}
\def \tl {\td \l}
\def \td {\tilde}
\def \Prod {\Pi}
\def \O {{\mathcal O}}
\def \Q {{\rm  Q}}
\def \D {\Delta}
\def \N {{\mathcal N}}
\def\tN{{\tilde N}}

\def \m {\mu}
\def \vs {\vec \s}
\def \ie {i.e.}

\def \cD {{\cal D}}

\def  \le  {\l_{\rm eff}}

\def \rS {{\rm S}}
\def\as{{\a}}
\newcommand{\bra}[1]{\mbox{$\langle #1 |$}}
\newcommand{\ket}[1]{\mbox{$| #1 \rangle$}}

\newcommand{\auth}{AUTHORS}

\def\thb{\bar{\theta}}
\def\Thb{\bar{\Theta}}
\def\barp{\bar{p}}
\def\barq{\bar{q}}
\def\barc{\bar{c}}
\def\bard{\bar{d}}
\def\e{\epsilon}

\def \bi{\bibitem}
\def \la {\label}

\def \l {\lambda}
\def\foot{\footnote}
\def \tl  {{\tilde \l}}
\def \sql {{\sqrt \l}}
\def \adss {$AdS_5 \times S^5$\ }
\newcommand{\rf}[1]{(\ref{#1})}
\def \ov {\over}

\def\th{\theta}
\def\Th{\Theta}
\def\vth{\vartheta}
\def\btheta{{\bar\theta}}
\def\ttheta{{{\tilde\theta}}}
\def\bttheta{{{\bar\ttheta}}}
\def\vth{\vartheta}

\def\ra{\rightarrow}
\def\N{{\cal N}}
\def\F{{\cal F}}
\def\uM{\underline{M}}
\def\uN{\underline{N}}
\def\uP{\underline{P}}
\def\cc{\circ}
\def\eqv{\equiv}

\def\ni{\noindent}
\def \ha{{1\ov 2}}
\def \bw {{\rm w}}

\def\r{{\rm r}}

\def\a{{\rm\bf a}}
\def\b{{\rm\bf b}}
\def\c{{\rm\bf c}}

\def\Y{{\rm Y}}
\def\X{{\rm X}}
\def\tY{\tilde{\rm Y}}
\def\tX{\tilde{\rm X}}
\def\dY{\dot{\rm Y}}
\def\dX{\dot{\rm X}}

\def \J {\mathcal{J}}
\def \del {\partial}

\def\dF{\dot{F}}
\def\dG{\dot{G}}
\def\df{\dot{f}}
\def \E {{\cal E}}

\def \S {{\cal S}}
\def \J {{\cal J}}

\def\ms{\mathcal{S}}
\def\mj{\mathcal{J}}
\def\soj{\fr{\ms}{\mj}}
\def \R {{\bf R}}
\def \om {\omega}
\def \tH {\widetilde H}
\def \bE {\bar E}
\def \x {{\cal X}}

 \def \bb {\bar \beta}
\def \W {{\cal E}}

\def \bi{\bibitem}
\def \la {\label}

\def \l {\lambda}
\def\foot{\footnote}
\def \tl  {{\tilde \l}}
\def \sql {{\sqrt \l}}
\def \sqtl {{\sqrt {\tilde \l}}}

\def \HH {{\rm E}}

\def \adss {$AdS_5 \times S^5$\ }

\def \D {\Delta}
\def \thet {\theta}
 \def \t {\tau}
 \def \p {\phi}
 \def \r {\rho}
 \def \rN {{\rm N}}
 \def\tw{{\tilde w}}
 \def\hJ{{J}}
 \def\hw{{w}}
 \def\hl{{\lambda}}
 \def\hth{{\theta}}
 \def\NN{{\cal N}}
 \def \bv {{ \bar w}}
\def \vn {{\vec n}}

\def \ov {\over}

\def \varpi {{\rm w}}
\def \OO {{\cal O}}

\begin{document}
\overfullrule=0pt
\parskip=2pt
\parindent=12pt
\headheight=0in \headsep=0in \topmargin=0in \oddsidemargin=0in

\vspace{ -3cm} \thispagestyle{empty} \vspace{-1cm}
\begin{flushright}
HUTP-05/A0046

\end{flushright}

\begin{center}

{\Large\bf $1/J^2$ corrections to BMN  energies from
\\   the quantum  long range Landau-Lifshitz model

 \vspace{0.01cm}
 }

 \vspace{.5cm} {
 J.A. Minahan$^{a,}$\footnote{On leave from Department of Theoretical Physics, Uppsala University}, A. Tirziu$^{c,}$\footnote{tirziu@mps.ohio-state.edu}
 and A.A.
 Tseytlin$^{d,c,}$\footnote{Also at
 Lebedev  Institute, Moscow.
 }}\\
 \vskip 0.3cm

{\em $^{a}$Jefferson Laboratory,
Harvard University\\
Cambridge, MA 02138 USA\\
\vskip 0.08cm $^{c}$Department of Physics, The Ohio State University,\\
Columbus, OH 43210, USA\\
\vskip 0.08cm $^{d}$  Blackett Laboratory, Imperial College,
London SW7 2BW, U.K. }

\end{center}

 \begin{abstract}
 In a previous paper (hep-th/0509071), it was shown that quantum $1/J$
 corrections to the BMN spectrum in an effective Landau-Lifshitz (LL) model match with the results
from the one-loop gauge theory, provided one chooses an
appropriate regularization.
  In this paper we continue this study for the conjectured Bethe ansatz
 for the long range spin chain
 representing perturbative \mbox{large $N$} $\NN=4$ Super Yang-Mills
in the $SU(2)$ sector, and the ``quantum string" Bethe ansatz for
its string dual.
 The comparison is carried out for corrections to BMN energies  up to
order $\tilde\lambda^3$ in the effective expansion parameter
$\tilde\lambda=\lambda/J^2$.  After determining the
``gauge-theory'' LL action to order
 $\tl^3$, which is accomplished indirectly by fixing the coefficients in the LL action
 so that the energies of circular
strings match with the energies found using the   Bethe ansatz, we
find
 perfect  agreement.  We interpret  this as further support for
an underlying integrability of the system.   We then consider the
``string-theory''  LL action which is a limit of the classical
string action representing fast string motion  on an $S^3$
subspace  of $S^5$ and
 compare the resulting  $\tl^3/J^2$  corrections to the prediction
 of the ``string'' Bethe ansatz.
As in the gauge case, we find   precise matching. This  indicates
 that the LL Hamiltonian supplemented  with a normal ordering
prescription  and $\zeta$-function regularization reproduces  the
full superstring result for the $1/J^2$ corrections,
 and also signifies
that  the  string Bethe  ansatz  does describe  the quantum BMN
string  spectrum to order $1/J^2$. We also comment on  using the
quantum LL approach to determine
 the non-analytic contributions in $\l$
that are behind the strong to weak coupling  interpolation between
the string and gauge results.

\end{abstract}
\newpage

\setcounter{equation}{0} \setcounter{footnote}{0}
\setcounter{section}{0}

\renewcommand{\theequation}{1.\arabic{equation}}
 \setcounter{equation}{0}

\section{Introduction}

Quantum corrections to semiclassical solutions of strings
propagating on $AdS_5\times S^5$ play an important part in the
investigation of AdS/CFT duality \ci{bmn,gkp,ft1}.    In
particular, the
 so-called three loop discrepancy
 between gauge and string predictions was first found when
  computing the leading $1/J$ correction\foot{For string
   computations, $1/\sqrt{\l}$ acts
   as an inverse string tension or as
  $\hbar$, which for semiclassical strings with large total $R$-charge $J$, can formally be traded with $1/J$. However,  it turns out that at higher orders
  of perturbation theory there are  additional genuine
 quantum corrections\cite{bt} which are reflected in the presence of terms
 non-analytic in $\lambda/J^2$.}  to the two-impurity BMN state \cite{callan}.
  The discrepancy was later found \cite{ss} to be present also for the semiclassical
   spinning string solutions \cite{ft2}.

The conclusion of \cite{callan}  (see also \cite{parn}) was the
result of
 a complicated calculation, and used  the contributions from the full set
 of world sheet fields, both bosonic and fermionic.
   Likewise, the quantum superstring   $1/J$ corrections were computed for the
   circular string solution of \cite{ft2},  again employing  the full set of
the bosonic and fermionic world-sheet fields \cite{ft3,fpt,ptt}.

In the gauge theory,  once one has found the dilatation operator,
 the fermionic excitations are not needed to compute $1/J$ corrections in the $SU(2)$
or the $SU(1,1)$ sectors, which are closed sectors containing no
fermion fields.\foot{In deriving the expression for the dilatation
operator one of course uses
 the full set of  bosonic and fermionic fields  of the SYM theory
 (for example, already at one loop,
  fermions contribute to the scalar self-energy diagrams).}  At the one-loop level,
that is to linear order in $\tl$, where $\tl$ is the effective
coupling $\tl=\l/J^2$, the corrections can be determined from the
corresponding Bethe ans\"atze for these sectors \cite{mz,bs}. At
higher loops, one can use the proposed long range Bethe ans\"atze
in \cite{ss,bds,s}.

Since fermions seem to play no role in the Bethe ansatz, we should
also be able to ignore them when computing $1/J$ corrections from
an effective action.
 In
 \cite{btz} and our previous paper \ci{mtt}
 this was shown to be the case when computing $1/J$
 corrections
 for the one-loop $SU(2)$ sector.
  The action used was the Landau-Lifshitz (LL) action, which is
  the effective action for the
  ferromagnetic Heisenberg spin chain in the continuum limit,
  with higher derivative counterterms to account for lattice effects.
  On the string side,
the counterpart of this action is the fast string limit around an
$S^3$ subspace of $S^5$ \cite{kru,krt,kt}. However, with only
bosonic $SU(2)$ sector modes being quantized, {\it i.e.} without
the rest of the superstring modes, including fermions, there are
infinities that need to be regularized. This can be accomplished
with a combination of normal ordering and $\zeta$-function
regularization.

A natural extension of \cite{mtt} is to carry out the computations
for higher orders in $\tl$.   In terms of the spin chain,
 this corresponds to going beyond nearest neighbor interactions, with
 order $\tl^n$ contributions coming from interactions between spins
 separated by up to $n$ sites.   In \cite{ss},
Serban and Staudacher (SS) first proposed an all-loop Bethe ansatz
that was based
 on the Inozemtsev spin chain \cite{inoz} and correctly reproduced the two and three loop predictions
 in \cite{bks}.  However, this Bethe ansatz  violated the
 BMN scaling at the 4 loop level, so a different ansatz was proposed by
 Beisert, Dippel and Staudacher  (BDS)  \cite{bds}
 that produces identical results as the SS ansatz to
  order $\tl^3$, but also preserves BMN scaling to all loops in the thermodynamic limit.

In order to compare results between a long range Bethe ansatz
calculation and an effective action calculation, we need to
 find the relevant  extension of the
  LL action.  The effective LL action to $\tl^2$ order
  was derived, both from the spin chain Hamiltonian and the fast string limit,
   in \cite{krt}.
   To go beyond $\tl^2$ order on the gauge side, the ``string'' LL
    action
   that follows from the fast string
    limit \cite{krt} can no longer be used,
     since the results on the gauge and string sides are known to
     disagree.

  In this paper, we are
able to find the ``gauge'' LL action to $\tl^3$ order indirectly,
by using the results from the SS/BDS Bethe ansatz for operators
that are dual to circular strings \cite{m}.
  We  construct the LL action
by including  all possible six derivative terms
 and varying their coefficients so that the
energies agree with the Bethe ansatz predictions. With an
effective action now available, we can then directly compute the
$1/J^2$ corrections to BMN states with $M$ impurities up to
$\tl^3$ order by quantizing the LL action (assuming  normal
ordering of the Hamiltonian  and using  a $\zeta$-function
regularization to remove further infinities).  Remarkably,
 comparing  the results to the ones  found directly from the
gauge theory Bethe ansatz, we find perfect agreement.

One can also do the same  on the string side, although in some
sense  the logic is in the reverse direction.
 Here one starts with the string effective
  action on $R\times S^3$ and takes the fast string limit,
  reducing the action to a ``string'' LL effective action.
  Results from the `string LL action can then be compared
  with results from the string ``quantum" Bethe ansatz of
  Arutyunov, Frolov and Staudacher
(AFS) \cite{afs}, which itself was originally derived by
``discretising'' the equations in \cite{kmmz} for general
classical string motion on $R\times S^3$.
 Again,  for the $1/J^2$ corrections for $M$-impurity
  BMN states we find, even more remarkably,
   perfect agreement up to $\tl^3$ order.

That these results match attests to the underlying integrability
of these systems.  Any system, integrable or otherwise, should be
describable by an effective action.
 However, the presence of a Bethe ansatz
means that all scattering amplitudes can be reduced to products of
two body scattering, the hallmark of integrability.  Our results
seem to indicate that the effective actions we use are consistent
with integrability, and that this integrability will be present at
the quantum level (or at least the first two orders),  even for
the string theory.

On the  gauge side, the LL action should be interpreted strictly
through its series expansion in $\tl$, since the 't Hooft coupling
$\l$ is the natural perturbative expansion parameter.
  However, for the string theory, $1/\sqrt{\l}$ is
the natural semiclassical  parameter and so for the string LL
action we are formally
 allowed to expand in $1/J$ while keeping $\tl$ fixed.
Hence,  when determining the  $1/J$ corrections, on the gauge
side, one should first expand in $\tl$ and then compute the
quantum corrections, while, on the string side, one should first
 compute the quantum corrections and then expand in $\tl$.
Because of the divergences that arise, the two procedures do not
commute and may lead to different results.
 In particular,  for the string
theory, this will lead to non-analytic terms in $\tl$ \cite{bt,sz}
and such
  non-analytic terms  should be included  only
 within the ``string'' interpretation of the LL computation.

We should stress  that the presence  of such non-analytic terms in
the near-BMN spectrum is non-trivial: on general grounds one
expects the  energy to have the following expansion \be\la{hah} E=
h_0(\tl) + {h_1(\tl) \ov J}  + {h_2(\tl) \ov J^2} + ... \ , \ee
and while  $h_0$ and $h_1$  are known  to have a regular expansion
in integer powers of $\tl$, the results of \ci{bt} suggest that
$h_2$ should contain non-analytic terms  with half-integer
 powers of $\tl$
starting with  $\tl^{5/2}$.  Below we will look for
 such non-analytic terms in the BMN spectrum
 using the quantum string LL approach,  with mixed results. Indeed,
  we do find half integer powers of $\tl$ in the
 $1/J^2$ corrections computing from string LL Hamiltonian,
but these come with logarithmic  divergences
 that needed to be regularized.
 Presumably,  for  the  full superstring
calculation the coefficients of these non-analytic terms will  be
finite, but we are   unable to unambiguously find these  finite
contributions using the string LL action.

\bigskip

This paper is organized as follows:  Section 2 describes the
structure of the  string and gauge LL actions,
 with the latter determined to $\tl^3$ order by comparing
 to results from the Bethe ansatz.  Section 3 is a review
 of the quantization procedure developed in \cite{mtt} relevant
  for BMN calculations, now applied to an LL action of more general structure.
  Sections 4 and 5 contain computations respectively for
  the $1/J$ and $1/J^2$ corrections.  Section 6 discusses
   non-analytic corrections and section 7 contains some concluding remarks.
Appendix A  describes how to fix  the structure of the gauge LL
action to $\tl^3$ order.
 Appendix B presents  computations of the energy of  $M$ impurity BMN states
 to $1/J^2$ order from  both the gauge and string Bethe ans\"atze.
Appendix C  discusses the structure of non-analytic terms in the
1-loop energy of a circular string solution.

\bigskip

\renewcommand{\theequation}{2.\arabic{equation}}
 \setcounter{equation}{0}

\section{Classical LL action to $\tl^3$ order}

Let us start by describing the structure of the LL action  viewed
as an effective action for low-energy excitations   on either the
string or gauge theory side (for a review see \ci{tse2,kt,mtt}).
On the gauge side  it is understood  in a  perturbative expansion
in $\tl= {\l \ov J^2}$,  and represents    the quantum effective
action for the low-energy spin wave modes of the spin chain
Hamiltonian equivalent to  the perturbative planar dilatation
operator in the $SU(2)$ sector\cite{mz}. On the string side it  is
a ``fast-string'' expansion of the classical string action in a
gauge \ci{krt,kt} where the density of  the momentum   of the
``fast'' collective coordinate  is constant.

It is known \cite{krt} that to ``2-loop'' or  $\tl^2$ order
 the LL actions obtained from the
string theory and gauge theory are the same. At ``3-loop'' or
$\tl^3$ order, however, they are  different. The coefficients in
the string LL action were  obtained  in  \ci{krt} while for  the
gauge-theory LL action one can fix them  by comparing the energy
of particular classical  solutions with the one obtained from the
spin chain Bethe ansatz. We shall discuss this in Appendix A.

As a result, one  may write  the  LL action  in the $SU(2)$ sector
as (we use the gauge where $t=\tau$;  $\ \partial_{1}  \equiv
\del_\sigma$)
\begin{equation}\label{aal}
S=J\int dt \int_{0}^{2\pi}\frac{d\sigma}{2\pi}\ L\,,
\end{equation}
 where
the  Lagrangian is
\begin{eqnarray}
L&=&  \vec C (n)   \cdot    \del_t \vn -\ \frac{1}{4}\vn\
\big(\sqrt{1-\tilde{\lambda}\partial_{1}^2}-1\big)
 \ \vn\ -
\frac{3\tilde{\lambda}^2}{128}(\partial_{1}\vn)^4\nonumber\\
&-&\frac{\tilde{\lambda}^3}{64}\left[\ \a \ (\partial_{1}\vn)^2
(\partial_{1}^2 \vn )^2\ +\ \b\ (\partial_{1}\vn
\partial_{1}^2 \vn)^2\  + \  \c \ (\partial_{1}\vn)^6\right]
+ O (\tl^4) \ .  \label{3loop}
\end{eqnarray}
Here $d C= \epsilon^{ijk} n_i d n_j \wedge d n_k$, i.e. $\vec C$
is a monopole potential on $S^2$. Also $\tl=\lambda/J^2$, where
$\lambda \sim 1/(\alpha')^2$. Here $\vn (t,\s)$ is a unit vector,
and we have included all terms which are quadratic in $\vn$.
   This exact quadratic  part follows from the string action
  \ci{krt} and also  from the coherent-state expectation value of the
  spin-spin part  of the dilatatiohttp://www.mozilla.org/products/firefox/start/n
  operator  on the gauge theory side \ci{rt}. It reproduces the BMN
  dispersion relation for small (``magnon'')  fluctuations near
  the BPS vacuum
  $\vec n = ( 0,0,1)$.

  The values of the ``3-loop'' coefficients in the string and
  gauge theory expressions for \rf{3loop}
   are:
\begin{equation}\la{oi}
\a _{s}=-\frac{7}{4}\ , \quad \quad \b_{s}=-\frac{25}{2}\ , \quad
\quad \c_{s}=\frac{13}{16}\ ,
\end{equation}
\begin{equation}\la{opim}
\a _{g}=-\frac{7}{4}\ , \quad \quad \b_{g}=-\frac{23}{2}\ , \quad
\quad \c_{g}=\frac{3}{4}\ .
\end{equation}
The string coefficients were found  in  \ci{krt}. The gauge
coefficients $\a $ and  $\c$ are fixed by comparing to the Bethe
ansatz results for the circular solution (see  Appendix A), while
the coefficients $\a $ and $\b$ can  be fixed by matching the
resulting $\tl^3/J$ correction to the  BMN energy to the
corresponding   gauge Bethe ansatz result \cite{bds,afs} (see
sect. 4).
\foot{In principle, to fix  the values  of the three coefficients
we could use, instead of the quantum $1/J$ BMN corrections,
  the classical folded
string solution, comparing its  LL energy to the Bethe ansatz
result of \cite{ss}.} We shall see in sect.
 5 that with these coefficients  the $\tl^3/J^2$ corrections
 also match, which provides
a strong consistency check.

 The difference between the string and gauge  values  of the
 coefficients $\b$ and $\c$  implies the difference between  the LL
 Lagrangians or the Hamiltonians
\begin{equation}\la{uup}
L_{s}-L_{g}= - ( H_s - H_g) = \frac{\tilde{\lambda}^3}
{64}\left[(\partial_{1}\vn\cdot\partial_{1}^2
\vn)^2-\frac{1}{16}(\partial_{1}\vn\cdot\partial_1\vn)^3\right]+
O(\tilde{\lambda}^4)
\end{equation}
This is a manifestation of the ``3-loop disagreement''
\ci{callan,ss}. Following  \ci{bt},  it can  explained  by
promoting the coefficients  $\b$ and $\c$  to functions of $\l$
such that for large $\l$ they approach the   string theory
values, while for small $\l$ they approach the gauge theory
values. Subleading terms in the string (strong-coupling) expansion
of $\b(\l)$ and $\c(\l)$, \be \la{gh} \b (\l) = \b_s +  {b_1\ov
\sqrt{\l} } + ..., \ \ \ \ \ \ \ \ \c (\l) = \c_s +  {c_1\ov
\sqrt{\l} } + ..., \ \ \ \ \ \ \l \gg 1 \ , \ee should come from
the part of the string  quantum corrections which are non-analytic
in $\tl$ \ci{bt}.
The quantum string effective action will then  have the structure
\rf{3loop} with  $\b(\l)$ and $\c(\l)$ as coefficients.
We shall return to the discussion of this
below.

As in  \cite{mtt},  let us  now rewrite  the LL Lagrangian
\rf{3loop} in terms of two independent fields. Solving the
constraint $|\vn|^2 =1$  as
 $n_3=\sqrt{1-n_{1}^2-n_{2}^2}$  we get the following  $SO(2)$
 invariant expression for the
 Lagrangian in terms of $n_1$ and $n_2$   ($a,b=1,2$; \  $n^2= n_a n_a$)
\begin{equation}
L=\  h^2(n)\
  \epsilon_{ab}\dot{n}_{a}n_{b} - H (n_1,n_2) \ , \ \ \ \ \ \ \ \ \ \
 h^2(n)=\frac{1-\sqrt{1-n^2}}{2n^2}={1 \ov 4 } + { 1 \ov 16} n^2 + ...\ , \ee
 \begin{eqnarray} \la{vector}
  H(n_1,n_2)&=  & \frac{1}{4} \ n_{a}\ (\sqrt{1-\tilde{\lambda}
  \partial_{1}^2}-1)\ n_{a}+
  \frac{1}{4}\ \sqrt{1-n^2}\ (\sqrt{1-\tilde{\lambda}
  \partial_{1}^2}-1)\  \sqrt{1-n^2} \nonumber\\
  && + \  \frac{3\tilde{\lambda}^2}{128} \ \bigg[n_{a}'^2+
  \frac{(n_{a}n_{a}')^2}{1-n^2}\bigg]^2 \nonumber\\
  && +\ \frac{\tilde{\lambda}^3}{64} \ \bigg\{
    \a   \bigg[n_{a}'^2+\frac{(n_{a}n_{a}')^2}
   {1-n^2}\bigg]\bigg[n_{a}''^2+
  \frac{\left(n_{a}n_{a}''(1-n^2)+n'^2-n_{a}^2
  n_{b}'^2+(n_{a}n_{a}')^2\right)^2}{(1-n^2)^{3}}\bigg]\nonumber\\
   &&\ \ \ \  \  +  \ \b \bigg[n_{a}'n_{a}''+\frac{n_{a}n_{a}'}
    {(1-n^2)^2}\bigg(n_{a}n_{a}''(1-n^2)+
  n'^2-n_{a}^2 n_{b}'^2+(n_{a}n_{a}')^2\bigg)\bigg]^2 \nonumber\\
   &&
\ \ \ \   \ + \  \c \bigg[n_{a}'^2+\frac{(n_{a}n_{a}')^2}
  {1-n^2}\bigg]^3
   \bigg\}\  +\  O (\tl^4) \la{laag}
   \ ,
\end{eqnarray}
where we use dot and prime for world-sheet time and space
derivatives.
 The function $h(n)$  has a regular
expansion near $n_a=0$,  and so  \rf{laag}
 may be  interpreted as a phase-space  Lagrangian
 with, say, $n_1$  being a coordinate  and $n_2$
being  related to  its momentum.

To simplify the quantization of the LL Lagrangian  near a
particular solution it is useful
 to put it  into the standard canonical form \ci{mtt}
 by doing the field
redefinition $n_a \to z_a $
\begin{equation}
z_{a}=h(n)\ n_{a} \ , \ \ \ \ \ \ \ \ \ \ n_{a}= 2\sqrt{1-z^2}\
z_{a} \ , \la{nz} \ee to obtain
\begin{equation}
L=\epsilon_{ab}\dot{z}_{a}z_{b}  -H(z_1,z_2) \ .  \label{vv}
\end{equation}
Having the Lagrangian in the standard form $L=p \dot q - H(p,q)$,
the quantization is straightforward: we  promote $z_a$ to
operators, impose the canonical commutation relation (cf.
\rf{aal})
\begin{equation}
[z_1(t,\s),z_2 (t,\s')]=   i J^{-1} \pi \delta (\s - \s')\,,
\label{cann}
\end{equation}
 and then decide how
to order the ``coordinate'' and ``momentum''  operators in
$H(z_1,z_2)$.


\renewcommand{\theequation}{3.\arabic{equation}}
 \setcounter{equation}{0}
\section{Quantization near  BPS vacuum:
corrections\\ to BMN spectrum from  LL  Hamiltonian}

As in \ci{mtt} our  aim will be to use the  LL action to compute
quantum
 $1/J$  and $1/J^2$ corrections to the
  BMN spectrum of fluctuations near
 the  BPS vacuum solution
\begin{equation}
\vec n=(0,0,1)\ , \ \ \
\end{equation}
representing the massless geodesic  in $R_t \times S^3$.
 The $1/J$
  corrections can be found
from the  Bethe ansatz on the spin chain \ci{mz,beisert} or from a
direct superstring quantization \ci{parn,callan}. As explained in
\ci{mtt}, the derivation  from the  LL action turns out to be
much simpler
 than the   string-theory derivation.
Here we  shall extend the method of \ci{mtt} to $\tl^3/J$
 and  $\tl^3/J^2$ orders.\foot{Unfortunately, the exact
 (all order in $\tl$)
 form of the $\vec n^4$ and $\vec n^6$ terms in the LL  action is not known,
 preventing us  from computing the $1/J$ and $1/J^2$ corrections to all
 orders in $\tl$.}
 The $1/J^2$  corrections to the BMN spectrum have not  yet been obtained from  a full superstring computation, and our LL approach
 provides a useful short-cut,  highlighting  several important
 issues that will also appear in the exact superstring
 approach.

Expanding  near this vacuum corresponds to expansion near $n_a=0$
in \rf{laag}
 or $z_a=0$ in
\rf{vv}. Observing that  the factor $J$ in front of  the LL action
\rf{aal}
 plays the role of the inverse Planck constant,
it is natural to rescale  $z_a$ as
\begin{equation}
z_{1}= \frac{1}{\sqrt{J}}\ f\ ,\ \ \ \  \quad z_{2}=
\frac{1}{\sqrt{J}}\ g\ , \la{fg}
\end{equation}
so that powers of $1/J$ will play the role of coupling constants
for the fluctuations in the  non-linear  LL Hamiltonian.
 Expanding the Hamiltonian in (\ref{laag}), (\ref{vv}) to  sixth  order in the fluctuation
 fields $f,g$  we get
\begin{equation}
S=\int dt \int_{0}^{2\pi}\frac{d\sigma}{2\pi}\ ( 2g \dot{f} - H )
\ , \    \ \ \ \ \ \ \ H = H_2 + H_4 +H_6 + ... \ , \la{qua} \ee
\be H_2 = f \ (\sqrt{1-\tilde{\lambda}\partial_{1}^2}-1) \ f\ + \
g \ (\sqrt{1-\tilde{\lambda}\partial_{1}^2}-1) \ g,  \label{quad}
\ee
\begin{eqnarray}
 H_4=&& { 1 \ov J} \bigg\{ (f^2+g^2) (\sqrt{1-\tilde
 {\lambda}\partial_{1}^2}-1)(f^2+g^2)-
  \ f(f^2+g^2)(\sqrt{1-\tilde{\lambda}
 \partial_{1}^2}-1)f\nonumber\\
 &&\ \  \ -\
  \ g(f^2+g^2)(\sqrt{1-\tilde{\lambda}\partial_{1}^2}-1)g
 \ +\frac{3\tilde{\lambda}^2}
 {8}(f'^2+g'^2)^2
 \nonumber\\
 &&\ \ \ +\ \frac{\tilde{\lambda}^3}{4}\left[\b (f'f''+g'
 g'')^2+\a (f'^2+g'^2)(f''^2+g''^2)\right]\bigg\}
  +  O ( {\tl^4\ov J})
\ , \label{quartic}
\end{eqnarray}
\begin{eqnarray}
 H_6= &&{ 1 \ov J^2} \bigg\{ \ \  \frac{1}{4}f(f^2+g^2)
 (\sqrt{1-\tilde{\lambda}\partial_{1}^2}-1)[f(f^2+g^2)]  \nonumber\\
&&\ \ \ \ \ \ +\   \frac{1}{4}g(f^2+g^2)
 (\sqrt{1-\tilde{\lambda}\partial_{1}^2}-1)[g(f^2+g^2)]
 \nonumber\\
 &&\ \ \ \ \ \ -\ \frac{1}{4}f(f^2+g^2)^2 (\sqrt{1-\tilde{\lambda}\partial_{1}^2}-1)f
 -\frac{1}{4}g(f^2+g^2)^2
 (\sqrt{1-\tilde{\lambda}\partial_{1}^2}-1)g\nonumber\\
 &&\ \ \ \ \ \ +\ \frac{3\tilde{\lambda}^2}{4}\bigg[2(f'^2+g'^2)(f f'+g
 g')^2-(f^2+g^2)(f'^2+g'^2)^2\bigg]\nonumber\\
 &&\ \ \ \ \ \ +\ \frac{\tilde{\lambda}^3}{2}\bigg[2 \c (f'^2+g'^2)^3- \b
 (f'f''+g' g'')\bigg((f^2+g^2)(f' f''+g' g'')\nonumber\\
 &&\ \ \ \ \ \ - \  (f'^2+g'^2)(f f'+g g')-2(f f'+g g')(f f''+g
 g'')\bigg)\nonumber\\
 &&\ \ \ \ \  \ \ + \ \a (f'^2+g'^2)\bigg(2(f'^2+
 g'^2)^2+(f f''+g g'')^2+3 (f'^2+g'^2)(f f''+g g'')\nonumber\\
&& \ \ \ \    \ \ - \ 2(f' f''+g' g'')(f f'+g
g')-(f^2+g^2)(f''^2+g''^2)\bigg)\nonumber\\
&&\ \ \ \ \ \ + \ \a (f f'+g g')^2 (f''^2+g''^2) \bigg] \bigg\}  +
O ( {\tl^4\ov J^2})
   \ .    \label{six}
\end{eqnarray}
Let us first consider the quadratic approximation. The linearized
equations of motion for the fluctuations are \be
\dot{f}=-(1-\sqrt{1-\tilde{\lambda}\partial_{1}^2})\ g\ , \quad\ \
\ \ \ \ \ \dot{g}=(1-\sqrt{1-\tilde{\lambda}\partial_{1}^2})\ f \
, \label{eqmotion} \ee and their  solution may be written as
\bea\label{fsol}
f(t,\sigma)&=&\frac{1}{2}\sum_{n=-\infty}^{\infty}(a_{n}e^{-i\omega_{n}t+in\sigma}+
a_{n}^{\dagger}e^{i\omega_{n}t-in\sigma})\ , \ \ \ \ \ \
\omega_{n} =  \sqrt{1+\tilde{\lambda}n^2}-1  \ ,\la{lo}\\
\label{gsol}
g(t,\sigma)&=&\frac{1}{2}\sum_{n=-\infty}^{\infty}(-ia_{n}e^{-i\omega_{n}t+in\sigma}+i
a_{n}^{\dagger}e^{i\omega_{n}t-in\sigma})\ , \la{loo} \eea for
real $f$ and $g$. Upon quantization  \rf{eqmotion} becomes the
equations of motion for the operators $f,g$
\begin{equation}\la{op}
\dot{f}=i[\bar H_2,f], \quad \quad \dot{g}=i[\bar H_2,g] \ , \ \ \
\ \ \ \bar H_2 \equiv   \int_{0}^{2\pi} \frac{d\sigma}{2\pi}\  H_2
\ ,
\end{equation}
provided we use the  canonical commutation relations in \rf{cann}
\begin{equation}\la{opi}
[f(t,\sigma),f(t,\sigma')]=0\  , \quad
[g(t,\sigma),g(t,\sigma')]=0\ , \ \ \ \
[f(t,\sigma),g(t,\sigma')]=i\pi \delta(\sigma-\sigma')\ .
\end{equation}
Then  the coefficients  in \rf{lo},\rf{loo}   satisfy
\begin{equation}\la{ui}
[a_{n},a_{m}^{\dagger}]=\delta_{n-m} \ ,
\end{equation}
so that   $a_{n}$ and $a_{n}^{\dagger}$ can be interpreted as
annihilation and creation operators, with the  vacuum state
$|0\rangle$ defined  by $a_n |0\rangle=0$, for all integer $n$.
 A general oscillator state is
\begin{equation}
|\Psi\rangle=\prod_{n=-\infty}^{\infty}
\frac{(a_{n}^{\dagger})^{k_{n}}}{\sqrt{k_{n}!}}|0\rangle \ .
\label{states}
\end{equation}
The  integrated  Hamiltonian $\bar H_2$  then becomes
\begin{equation}\label{ham2}
\bar H_2 = \sum_{n=-\infty}^{\infty}\ \omega_{n}\ a_{n}^{\dagger}
a_{n} \ , \ee where we have used the normal ordering to ensure
that the vacuum energy is   zero,  since the BMN vacuum is  a BPS
state in both gauge theory and string theory.

One also needs to impose the  extra constraint that the total
$\sigma$-momentum is zero  \cite{mtt}.
For  physical oscillator states we get \be \la{cons}
\sum_{n=-\infty}^{\infty}n a_{n}^{\dagger}a_{n} |\Psi\rangle =0\ ,
\ \ \ \ \ \ \ \ \sum_{n=-\infty}^{\infty}n k_n =0 \ . \ee Below we
shall  consider the
 ``$M$-impurity'' states as oscillator states with $k_n=1$:
 \begin{equation}
|M\rangle = a_{n_{1}}^{\dagger}...a_{n_{M}}^{\dagger}|0\rangle \ ,
\end{equation}
 where for  simplicity we  shall assume that  all $n_j$  are
 different (generalization to  states
 with several equal $n_j$ is straightforward,   at
 least for  $1/J$ corrections).
Then the  zero-momentum condition \rf{cons}   gives
\begin{equation}\label{momcond}
\sum_{j=1}^{M}n_{j}=0 \ ,
\end{equation}
and the  leading  term in the energy of an
 $M$-impurity state takes  the familiar  form  \ci{mets,bmn}
\begin{equation}
E^{(0)} = \langle  M|\bar H_2 |M\rangle=\sum_{j=1}^{M}(\sqrt
{1+\tilde{\lambda}n_{j}^2}-1)\ . \la{enn}
\end{equation}
It is useful to make  a  comment on the choice of parameters.
 In the
LL  approach we use $J= J_1 + J_2 $ as a natural total angular
momentum, corresponding to a ``fast'' collective coordinate. Here
$M$ is  a characteristic of a particular state,  while
  it is
  $J$ that enters into the
background-independent form  of the LL action  \rf{aal}.
 This is
in line  with gauge/spin chain intuition,  where the use  of total
$J$ or spin chain length as the   state-independent parameter is
natural. At the same time, on the string side, when   expanding
near a BPS  state, {\it i.e.} a massless geodesic with spin
$J_1$,  one builds up $J_2$ from quantum excitations,   and here
it is natural to use
 $J'=J_1$ and $M=J_2$ as the basic parameters of, respectively,
  the vacuum and the state.
  Thus,  compared to generic states in the $SU(2)$ sector that carry
spins $(J_1,J_2)$ with $J= J_1 + J_2$, here  we have \ci{mtt}
\begin{equation}\la{er}
J_1=J-M\ ,\ \ \ \ \ \  \quad J_{2}=M \ .
\end{equation}
The corresponding gauge-theory states are Tr$(\Phi_1
^{J_{1}}\Phi_2 ^{J_{2}})+ ...$, and $J$ plays the role  of the
length of the spin chain and $M$ is the number of magnons. On the
string side, the LL approach is adapted
 to  semiclassical solutions
for which   $J_1$ is of order $J_2$ rather than to  near-BMN
states which are small fluctuations near the vacuum  and for which
$J_1 \gg J_2$. In describing BMN states in the LL approach one has
an ``unnatural'' choice of parameters:
 $\tilde{\lambda}  \equiv \lambda/J^2$, not
 the usual BMN effective coupling
 $\lambda'  \equiv \lambda/{{J}'}^2$.
In the LL description the  BMN   energy $E$  starts with $J$ to
which we add terms of order $\tilde{\lambda}$, i.e.
  $E= J + E^{(0)} +  O({1 \ov J}) = J  +  \sum_{j=1}^{M}(\sqrt
{1 + {\tl}n_{j}^2} -1 ) +  O({1 \ov J}),  $
 while  the equivalent
  string theory expression is   $  E= J' +  \sum_{j=1}^{M}\sqrt
{1 + {\lambda'}n_{j}^2} +  O({1 \ov J'}) $.

\renewcommand{\theequation}{4.\arabic{equation}}
 \setcounter{equation}{0}

\section{$1/J$ corrections to the BMN spectrum}

Let us now generalize the  computation of the $1/J$ corrections to
the energy \rf{enn} in \ci{mtt} to  $\tl^3$ order. To compute  the
$1/J$ correction  to the energy of an $M$-impurity state
 one needs to include  the quartic term in the Hamiltonian \rf{quartic}
  integrated over $\s$, i.e.
 $\bar H_4\equiv  \int_{0}^{2\pi}\frac{d\sigma}{2\pi} H_4$,
 and use the standard  quantum mechanical perturbation theory.
Written in  terms of the creation and annihilation operators,
$\bar H_4$ is found to be
\begin{eqnarray}\label{ham4}
\bar H_4 &=&\frac{1}{J}\sum_{n,m,k,l} \bigg[ c_{nmk}
+\frac{3\tilde{\lambda}^2}{32}n m k
l+\frac{\tilde{\lambda}^3}{16}\left( \a \ n m k^2 l^2 -\b\  n m^2
k l^2\right) + O ({\tl^4})
\bigg]\nonumber\\
&&\times \bigg(a_{n}a_{m}^{\dagger}a_{k}a_{l}^{\dagger}+
a_{n}a_{m}^{\dagger}a_{-k}^{\dagger}a_{-l }+a_{n}^{\dagger}a_{m}
a_{-k}a_{-l}^{\dagger}+ a_{n}^{\dagger}a_{m}a_{k}^{\dagger}a_{l}
\bigg)   \delta_{n-m+k-l} \ , \la{hea}\end{eqnarray} where
\begin{equation}
c_{nmk}\equiv \frac{1}{4}\sqrt{1+\tilde{\lambda}(n-m)^2}-
\frac{1}{8}\sqrt{1+\tilde{\lambda}n^2}
-\frac{1}{8}\sqrt{1+\tilde{\lambda}(n-m+k)^2}\ .
\end{equation}
In the expression for  the interacting Hamiltonian
 we have dropped  the time
dependent phases ($e^{-i \w_n t}$) since they can be removed by a
unitary transformation with the quadratic Hamiltonian
$\bar{H}_{2}$. Here and in what follows the summations over $n,m$,
{\it etc.}, are from $-\infty$ to $\infty$.

As discussed  in \cite{mtt}, to obtain the results consistent with
both  the gauge-theory  spin chain and the string-theory
expressions one should use a normal ordering prescription for
$\bar H_4$. Doing so we get \be \bar H_4= \frac{1}{J} \sum_{n,m}
\
 h_{nm} \ a_{n}^{\dagger}a_{m}^{\dagger}a_{n}a_{m}\ , \la{hi} \ee
\bea h_{nm}= &&1+\sqrt{1+\tilde{\lambda}(n-m)^2}-
2\sqrt{1+\tilde{\lambda}n^2} \nonumber \\
&& \ \ +\  \frac{3\tilde{\lambda}^2}{4}n^2
m^2+\frac{\tilde{\lambda}^3}{16 } n^2 m^2 [  2\a ( n+ m)^2 +  \b
(n-m)^2  ]
 +   O ({\tl^4})   \ . \la{hii}
\eea Then  the leading  correction to the energy  \rf{enn} of an
$M$-impurity state is given by
\begin{eqnarray}
&& \langle  M|\bar H_4 |M\rangle  =
\frac{1}{J}\bigg\{M^2-2M+\sum_{i,j=1}^{M}\sqrt{1+\tilde{\lambda}(n_i-n_{j})^2}-
2(M-1)\sum_{i=1}^{M}\sqrt{1+\tilde{\lambda}n_i^2}\nonumber\\
&&\qquad \qquad \qquad \qquad \ \ \ \ \
 \ \ +\ \frac{3\tilde{\lambda}^2}{4}\sum_{i=1}^{M}
n_i^4\ \ +\ \ \frac{\tilde{\lambda}^3}{16}\bigg[\ \ - 8 \a
\sum_{i=1}^{M}n_i^6
\nonumber\\
&&\qquad \qquad \qquad \ \ \ \ \ \   + \ \sum_{i,j=1}^{M}n_i^2
n_{j}^2 \big[\   2\a \ ( n_i+ n_j )^2
 + \  \b\  (n_i -n_j )^2\big]  \bigg]  + O(\tl^4)
 \bigg\}\ . \la{hw}
\end{eqnarray}
Expanding in $\tl $ gives for the $1/J$ correction to the energy
\begin{eqnarray}\la{ooo}
&&E^{(1)}= \langle  M|\bar H_4 |M\rangle= { 1 \ov J} \bigg\{ \tl
\sum_{i=1}^{M}n_i^2- \tilde{\lambda}^2  \sum_{i=1}^{M}n_i^4 +  { 1
\ov 8} ( 1 - 4 \a ) \tl^3 \sum_{i=1}^{M}n_i^6 \nonumber\\   && \ \
+ \ \frac{\tilde{\lambda}^3}{16}
 \sum_{i,j=1}^{M}  n_{i}^2n_{j}^2 \bigg[
   (2 \a + \b  + 15)  (n^2_i + n^2_j )  + 2 ( 2 \a - \b -10)
    n_i n_j \bigg]
 +O(\tilde{\lambda}^4)   \bigg\}
\end{eqnarray}
Plugging in the  string-theory and gauge-theory coefficients $\a ,
\b$ in \rf{oi},\rf{opim} we conclude that this expression  is in
precise  agreement
 with the full string theory
 computation in \ci{callan,swanson} and  with the result
 found  using  the gauge and  string Bethe  ans\"atze
in \ci{afs}, expanded up to $\tl^4/J$ order.
 This agreement confirms, in particular,
   the values of the coefficients
 $\a $ and $\b$ in the gauge-theory LL  action given in  \rf{opim}
 (see also Appendix A).

Let us recall  again  that in
  comparing  with the near-BMN results of \ci{afs,swanson},
one should  note that  $J$ as  defined there is  $ J_1 $ in the
$SU(2)$ sector notation. To compare with our results, one may
define $J'\equiv J_1= J-M$ and $\l'= {\l\ov J'^2}$; the
expressions  of \ci{afs,swanson} should have
 $(J,\tl)$ replaced with $(J',\l')$ and
then   re-expressed in terms of the  parameters $(J,\tl)$ which
are    natural in the present LL approach.

The difference between  the order $\tl^3/J$
 string and gauge theory corrections to the BMN energy
 is because of the difference of the values
 of the coefficient  $\b$:  $\b_s- \b_g= -1$.  The energy difference
  is thus  \ci{bt}
 \begin{equation}\la{kot}
E^{(1)}_{s}-E^{(1)}_{g}=  -\frac{\tl^3}{16 J} \ \sum_{i,j=1}^{M}
n_i^2 n_j^2 ( n_i-n_j)^2+O(\frac{\tl^4}{J})\ .
\end{equation}
 We shall return to the discussion of this difference
 in sect. 6.

In the string case the last double-sum term in  \rf{ooo} is $ -
\frac{\tilde{\lambda}^3}{16}
 \sum_{i,j=1}^{M}  n_{i}^2n_{j}^2  (n_{i} + n_{j})^2 $
 while in the  gauge case
it is $ - \frac{\tilde{\lambda}^3}{4}
 \sum_{i,j=1}^{M}  n_{i}^3n_{j}^3 . $
As discussed in Appendix B, the gauge-theory expression in
\rf{ooo} has a simple spin-chain generalization to all orders in
$\tl$ implied by the gauge Bethe ansatz \ci{bds,afs}\foot{This is
equivalent to eq. (3.8) in \ci{afs} after observing that $J$ there
is $J-M$ here and $\l'$ there is $\l \ov ( J-M)^2$ here. As
always, we assume  the condition \rf{momcond}.}
\begin{equation}
E^{(1)}_g =\frac{\tl}{J} \bigg( \sum_{ i=1}^M\frac{n_i^2}{ 1 + \tl
n^2_i } - \sum_{i, j=1}^M\frac{n_in_j}{ \sqrt{ 1 + \tl n^2_i } \
\sqrt{ 1 + \tl n^2_i }  }   \bigg)\,.
\end{equation}

\renewcommand{\theequation}{5.\arabic{equation}}
 \setcounter{equation}{0}

\section{$1/J^2$   corrections to BMN spectrum}

To find $1/J^2$ corrections we follow the method in \cite{mtt}
where order $\tl/J^2$ terms were 
computed.  We need to combine   the second order perturbation
theory correction for the quartic Hamiltonian \rf{quartic} with
the first order perturbation theory correction
 for the sixth
order Hamiltonian in \rf{six}. The regularization issues were
discussed in detail in \ci{mtt}: to match string/gauge results  we
 should use the normal-ordered form  of the Hamiltonians and 
apply $\zeta$-function  regularization for intermediate-state
sums. We shall also need to add a local higher-derivative
``counterterm'' which (on gauge side) is a lattice correction to
the continuum limit of the LL action (see \ci{mtt} and below).

\bigskip
\subsection{Second-order perturbation (``exchange'')   contribution}
\bigskip

Starting with the  quartic Hamiltonian \rf{hea} we  need to
compute
\begin{equation}\la{secc}
 \langle  M| (\bar H_4)^{(2)} |M\rangle=
  \sum_{M\neq
M'}\frac{\langle  M|\bar H_4 |M'\rangle \langle  M'| \bar H_4
|M\rangle}{E_M-E_{M'}}  \  ,
\end{equation}
where $|M'\rangle $ is any possible intermediate state, and
 $|M\rangle=a^\dagger _{n_{1}} ... a^\dagger_{n_{M}}|0\rangle$.
 Since $\bar H_4$ in \rf{hi} contains only terms of  the
form $(a^{\dagger})^2a^2$, the only non-trivial intermediate
states can be the $M'=M$ -particle states 
 of the form $ a^\dagger
_{n'_{1}} ... a^\dagger_{n'_{M}}|0\rangle  $.
 Then in order for
the matrix element $\langle  0 | a_{n_{1}} ... a_{n_{M}}
 |\bar H_4 | a^\dagger _{n'_{1}} ...
a^\dagger_{n'_{M}}|0\rangle $ to be non-zero, there should be a
$j$ and $k$ such that   $n'_j=n_j+q$ and $n'_k= n_k-q$, with all
other $n_i'=n_i$, $i\ne j,k$.  In order for $|M\rangle$ to be
distinct from $|M'\rangle$, we require that $0\ne q\ne n_k-n_j$.
With these conditions, we then find that if $n_k\ne n_j$
\begin{eqnarray} \la{uio}
\HH_{4}^{(1)}\equiv &&\langle  M|\bar H_4 |M'\rangle
=\frac{1}{J}\bigg\{  {2}\sqrt{1+\tilde{\lambda}q^2}
+2\sqrt{1+\tilde{\lambda}(n_{k}-n_{j}-q)^2} \nonumber\\ &&\ \ \ \-
\ \sqrt{1+\tilde{\lambda}n_{k}^2}-
\sqrt{1+\tilde{\lambda}n_{j}^2}-\sqrt{1+
\tilde{\lambda}(q+n_{j})^2} - \sqrt{1+\tilde{\lambda}(n_{k}-q)^2}
\nonumber\\
&& \ \ \ \  +  \ \frac{3\tilde{\lambda}^2}{2}n_{k}n_{j}(q+n_{j})
(n_{k}-q)\bigg[1+\frac{\tilde{\lambda}}{12}\bigg(2\a
(n_{j}+n_{k})^2
\nonumber\\
&& \ \ \ \ \ \ \ \ \ \   + \ \b[(n_{k}-n_{j})^2-2q
(n_{k}-n_{j}-q)] \bigg)\bigg]\bigg\},\
\end{eqnarray}
 where $n_j+q$ and  $n_k-q$ are not equal to
one of the other $n_l$'s. The energy difference in \rf{secc} is
\begin{equation}\la{kk}
W_{1}\equiv
E_M-E_{M'}=\sqrt{1+\tilde{\lambda}n_{j}^2}+\sqrt{1+\tilde{\lambda}n_{k}^2}
-\sqrt{1+\tilde{\lambda}(n_{j}+q)^2}-\sqrt{1
+\tilde{\lambda}(n_{k}-q)^2}\ .
\end{equation}
 If $n_j+q=n_l$,
and so $|M'\rangle$ has two impurities with the same momenta, then
the matrix element is
\begin{eqnarray}\la{oop}
\HH_{4}^{(2)}\equiv &&\langle  M|\bar H_4 |M'\rangle
=\frac{\sqrt{2}}{J}\bigg\{
2\sqrt{1+\tilde{\lambda}(n_{l}-n_{j})^2}
+2\sqrt{1+\tilde{\lambda}(n_{k}-n_{l})^2}
\nonumber\\
&& \ \ \ \ \ - \ \sqrt{1+\tilde{\lambda}n_{k}^2}-
\sqrt{1+\tilde{\lambda}n_{j}^2}-\sqrt{1+\tilde{\lambda}n_{l}^2} -
\sqrt{1+\tilde{\lambda}(n_{k}-n_{l}+n_{j})^2}
\nonumber\\
&&\ \ \ \ \ \ \ \ \ \ \ \ + \
\frac{3\tilde{\lambda}^2}{2}n_{k}n_{j}n_{l}
(n_{k}+n_{j}-n_{l})\bigg(1+ \frac{\tilde{\lambda}}{12}\bigg[2\a
(n_{j}+n_{k})^2
\nonumber\\
&& \ \ \ \ \ \ \ \ \ \ \ \ \ \ + \
\b[(n_{k}-n_{j})^2-2(n_{l}-n_{j}) (n_{k}-n_{l})]
 \bigg] \bigg) \bigg\},\
\end{eqnarray}
and the energy difference is
\begin{equation}
W_{2}\equiv
E_M-E_{M'}=\sqrt{1+\tilde{\lambda}n_{j}^2}+\sqrt{1+\tilde{\lambda}n_{k}^2}
-\sqrt{1+\tilde{\lambda}n_{l}^2}-\sqrt{1+
\tilde{\lambda}(n_{j}+n_{k}-n_{l})^2} .
\end{equation}
Then  the ``exchange'' contribution is given by
\begin{equation}\label{H42sum}
\langle  M| (\bar H_4)^{(2)} |M\rangle
=\frac{1}{4}\sum^M_{j,k\atop j \ne k}\bigg[
\sum_{q=-\infty\atop0\ne q\ne n_k-n_j}^\infty
\frac{(\HH_{4}^{(1)})^2}{W_{1}}+\sum^M_{l\ne j\atop l\ne
k}\frac{(\HH_{4}^{(2)})^2}{W_{2}} \  \bigg] \ .
\end{equation}
Interpreted as a string-theory  expression (i.e. non-perturbative
in $\tl$, with sums done before the expansion in $\tl$), the sum
over the ``virtual'' momentum $q$  produces, as we shall see in
section 6, a contribution which  is non-analytic in $\tl$. This is
a novel phenomenon \ci{bt}, which was  absent at order $1/J$.

Let us first  ignore such contributions and expand in $\tl$ before
doing the sum over $q$.  (This is the procedure which is in any
case appropriate for the gauge-theory.) This  will give us
 terms which are
 analytic  in $\tilde{\lambda}$.
 It was shown in \cite{mtt} that the quantum
LL  Hamiltonian and the gauge theory Bethe ansatz give the same
$\tilde{\lambda}/J^2$ correction (the  exact superstring result is
not yet available). Our aim here is to extend  the computation of
\ci{mtt} to  the  orders $\tilde{\lambda}^2/J^2$ and
$\tilde{\lambda}^3/J^2$ and to  show that the results match
 the gauge and string  Bethe ansatz results found
in Appendix B.

For the part of $\tilde{\lambda}^2/J^2$ correction coming from the
``exchange'' contribution \rf{secc},\rf{H42sum} we obtain
\begin{eqnarray}
\frac{\tilde{\lambda}^2}{J^2}\bigg[\sum_{k,j=1\atop k \ne j}^{M}
\frac{n_{j}^3 n_{k}^3}{(n_{k}-n_{j})^2}+ \frac{9}{8} \big( M-4
\big) \bigg(\sum_{i=1}^{M}n_{i}^2\bigg)^2+  {1 \ov 8} \big( M^2
-8M + 20 \big)\sum_{i=1}^{M}n_{i}^4\bigg] \ . \label{2sec}
\end{eqnarray}
We  see that the pole-type contribution  $\sim {1\ov
(n_{k}-n_{j})^2}$ matches the corresponding  term in the
correction to the BMN  energy (\ref{E2final1}) computed from the
Bethe ansatz. To compare  other ``local'' terms we must add the
 ``contact''  contribution  coming from the expectation value of the
 sixth-order  Hamiltonian and also add a local counterterm \ci{mtt}
 contribution (see below).
Note that at orders $\tl$ and $\tl^2$ the string and gauge  Bethe
ans\"atze produce the same expressions, so they both agree with
what we find from the LL approach.

\bigskip

Next, let us  consider the order $\tilde{\lambda}^3/J^2$ term. We
first concentrate on finding the pole term
  $\sim {1\ov (n_{k}-n_{j})^2}$
in the first line
 of (\ref{E2final1}). It can come only  from the first
term in (\ref{H42sum}), where the order $\tilde{\lambda}^3/J^2$
part is
\begin{eqnarray}
\frac{(\HH_{4}^{(1)})^2}{W_{1}} \to  { \tl^3 \ov 16J^2} \frac{1}{
v_{kj} } \bigg[A_{1}+ v_{kj} \bigg( A_{2}+  v_{kj} [A_{3}- v_{kj}
(A_{4}-2v_{kj})] \bigg)\bigg]\ ,
\end{eqnarray}
\be  v_{kj} \equiv q(n_{k}-n_{j}-q)  \ . \la{vee} \ee Here
$A_{1},A_{2},A_{3},A_{4}$ are polynomials in $n_{j}$  which  do
not depend on $q$. The pole  term comes from (see also \ci{mtt})
\begin{eqnarray}
&&\frac{1}{4}\sum_{j,k\atop j\ne k}\sum_{q}
\frac{(\HH_{4}^{(1)})^2}{W_{1}} \rightarrow\sum_{k,j=1\atop k \ne
j}^{M}\sum_{q=-\infty\atop0\ne q\ne n_k-n_j}^\infty
\frac{A_{1}}{64q(n_{k}-n_{j}-q)}=-\frac{1}{32}\sum_{k,j=1\atop k
\ne j}^{M}\frac{A_{1}}{(n_{k}-n_{j})^2}\nonumber\\
&&\ \ =\frac{1}{4}\b \sum_{k,j=1\atop k \ne j}^{M}n_{j}^3 n_{k}^3
-\frac{1}{4}\sum_{k,j=1\atop k \ne
j}^{M}\frac{n_{j}^2n_{k}^2}{(n_{k}-n_{j})^2}\bigg[3n_{j}^4+3n_{k}^4-18
n_{j}^3 n_{k}-18 n_{j}n_{k}^3+19 n_{j}^2 n_{k}^2\nonumber\\
&&\qquad \qquad \qquad \qquad\ \qquad \qquad \ \ \ \ \ \ \ \ \ \ \
\ \ -\ 2 \a \ n_{j}n_{k}(n_{j}+n_{k})^2\bigg]\ .
\end{eqnarray}
We observe that the pole  term depends only on  $\a$, which
 is  the same for the gauge and string LL actions.
Setting  $\a =-7/4$ as in (\ref{oi}), (\ref{opim})  we get for the
pole term
\begin{equation}
-\frac{3\tilde{\lambda}^3}{4J^2} \sum_{k,j=1\atop k \ne j}^{M}
\frac{n_{j}^4n_{k}^4}{(n_{k}-n_{j})^2} \  .
\end{equation}
This  indeed matches the expression for the
 corresponding pole  term in  both the gauge and string Bethe ansatz
results (\ref{E2final1})  and  (\ref{E2final2}).

Computing the sums in (\ref{H42sum}) with  all other ``local''
terms included we find the following total  result  for the
$\tl^3/J^2$ contribution from the second order perturbation theory
correction\foot{As in \ci{mtt} here we used the  $\zeta$-function
regularization, i.e. set $\sum_{n=-\infty}^\infty n^s=0$ for
$s=0,1,2,....$.}
\begin{eqnarray}
 E^{(2)}_{3 \ exch.} =&& \frac{\tilde{\lambda}^3}{48 J^2}\bigg[(113+56 \a
)\bigg(\sum_{j=1}^M n_{j}^2\bigg)^3-6(116 +8
\b-5M)\bigg(\sum_{j=1}^M n_{j}^3\bigg)^2\nonumber\\
&&+\ (304 +136 \a +36 \b+96 M-3 M^2) \sum_{j=1}^M n_{j}^6
\label{3loopsec}
\\
&& +\ (426
 -192 \a +24 \b-135 M) \sum_{i=1}^M n_{i}^4 \sum_{j=1}^M
n_{j}^2  + \ 12(8 \a +11)\sum_{i\ne
j}^M\frac{n_i^4n_j^4}{(n_i-n_j)^2}\bigg]
 \nonumber .
\end{eqnarray}

\bigskip
\subsection{Sixth-order ``contact''   contribution}
\bigskip
Let us now compute the  contribution coming from the expectation
value of the  sixth order term in the LL  Hamiltonian (\ref{six}).
After a lengthy computation we obtain the following expression for
the normal ordered form for this term\foot{In this subsection  the
only
 regularization we use is the assumption that the
sixth order term in the Hamiltonian is normal ordered.}
\begin{equation}
\bar{H}_{6} = \frac{1}{J^2}\sum_{n,m,k}\  h_{nmk} \
a_{n}^{\dagger}a_{m}^{\dagger}a_{k}^{\dagger}a_{n}a_{m}a_{k}\ ,
\la{hhh}
\end{equation}
\begin{eqnarray}\la{cof}
&&h_{nmk}=\frac{1}{4}\bigg[\sqrt{1+\tilde{\lambda}(n-m+k)^2}+
\sqrt{1+\tilde{\lambda}(n-m-k)^2}-2\sqrt{1+\tilde{\lambda}n^2}\ \bigg]\nonumber\\
&&\  \ \ \ \  \ \ \   + \ \frac{1}{2}n^2 m k \bigg\{ - 9 \tilde{\lambda}^2 \\
&&\ \ \ \  \  \ \ \    + \ \tilde{\lambda}^3  \bigg[(9 \a + 12 \c)
\  m k - (\a + \b)  n^2 +  (\b - 14 \ \a) \ n m  \bigg] \bigg\} +
O(\tl^4)\,  . \nonumber
\end{eqnarray}
Expanding the square roots in $h_{nmk}$  in  powers of
$\tilde{\lambda}$ we can find contributions to the $1/J^2$
correction to order $\tilde{\lambda}^3$. At order
$\tilde{\lambda}$ we already computed  the resulting expectation
value  in \cite{mtt}. At order $\tilde{\lambda}^2$ we have
\begin{eqnarray}
h^{(2)}_{mnk}= -\frac{\tilde{\lambda}^2}{8 J^2} (n^4+9 n^2 m^2 -4
n^3 m+30 n^2 m k) \ .
\end{eqnarray}
For the expectation vales in an $M$-impurity state   we get
\begin{equation}
\langle M|\sum_{n,m,k}n^4
a_{n}^{\dagger}a_{m}^{\dagger}a_{k}^{\dagger}a_{n}a_{m}a_{k}|M\rangle=(M-1)(M
-2) \sum_{j=1}^M n_{j}^4 \ ,
\end{equation}
\begin{equation}
\langle M|\sum_{n,m,k}n^2 m^2
a_{n}^{\dagger}a_{m}^{\dagger}a_{k}^{\dagger}a_{n}a_{m}a_{k}|M\rangle=
(M-2)\bigg(\sum_{j=1}^M n_{j}^2\bigg)^2- (M-2) \sum_{j=1}^M
n_{j}^4\ ,
\end{equation}
\begin{equation}
\langle M|\sum_{n,m,k}n^3 m \
a_{n}^{\dagger}a_{m}^{\dagger}a_{k}^{\dagger}a_{n}a_{m}a_{k}|M\rangle=
-(M-2)\sum_{j=1}^M n_{j}^4\ ,\
\end{equation}
\begin{equation}
\langle M|\sum_{n,m,k}n^2 m k \
a_{n}^{\dagger}a_{m}^{\dagger}a_{k}^{\dagger}a_{n}a_{m}a_{k}|M\rangle=
-\bigg(\sum_{j=1}^M n_{j}^2\bigg)^2+2 \sum_{j=1}^M n_{j}^4\ .
\end{equation}
Then the $\tl^2$ contribution is  found to be
\begin{equation}
E^{(2)}_{2 \ cont.}= -\frac{\tilde{\lambda}^2}{8J^2}\bigg[
\big({9M}- 48\big) \bigg(\sum_{j=1}^M n_{j}^2 \bigg)^2 +
\big({M^2}-8M+72 \big) \sum_{j=1}^M n_{j}^4\bigg] \  .
\end{equation}
Adding it to the corresponding exchange term (\ref{2sec}) we
observe that all $M$-dependent coefficients cancel and the result
is
\begin{equation}
E^{(2)}_{2\ exch.+cont.} = \frac{\tilde{\lambda}^2}{J^2}\bigg[
\frac{3}{2} \bigg(\sum_{j=1}^M n_{j}^2
\bigg)^2-\frac{13}{2}\sum_{j=1}^M n_{j}^4+\sum_{k,j=1\atop k \ne
j}^{M} \frac{n_{j}^3 n_{k}^3}{(n_{k}-n_{j})^2}\bigg]\ .
\label{2tot}
\end{equation}
The $\tl^3$ term in $h_{mnk}$ is
\begin{eqnarray}
&& h^{(3)}_{mnk}= \frac{\tilde{\lambda}^3}{16J^2}  \bigg[n^6+45
n^4
m^2- (15 + 8 \ \a +8 \ \b)\ n^4 m k -  10 n^3 m^3\nonumber\\
&&-\ (60+112 \ \a - 8  \ \b) \ n^3 m^2 k  -6  n^5 m +(45 +72 \ \a
+ 96 \ \c) \ n^2 m^2 k^2 \bigg]   \   . \label{3loopsix}
\end{eqnarray}
For $M$-impurity states we get
\begin{equation}
\langle M|\sum_{n,m,k}n^6
a_{n}^{\dagger}a_{m}^{\dagger}a_{k}^{\dagger}a_{n}a_{m}a_{k}|M\rangle=(M-2)(M-1)
\sum_{j=1}^M n_{j}^6 \ ,
\end{equation}
\begin{equation}
\langle M|\sum_{n,m,k}n^4 m^2
a_{n}^{\dagger}a_{m}^{\dagger}a_{k}^{\dagger}a_{n}a_{m}a_{k}|M\rangle=(M-2)
\sum_{j=1}^M n_{j}^4  \sum_{k=1}^M n_{k}^2-(M-2)\sum_{j=1}^M
n_{j}^6 \ ,
\end{equation}
\begin{equation}
\langle M|\sum_{n,m,k}n^4 m k \
a_{n}^{\dagger}a_{m}^{\dagger}a_{k}^{\dagger}a_{n}a_{m}a_{k}|M\rangle=-
\sum_{j=1}^M n_{j}^4  \sum_{k=1}^M n_{k}^2+2\sum_{j=1}^M n_{j}^6\
,
\end{equation}
\begin{equation}
\langle M|\sum_{n,m,k}n^3 m^2 k \
a_{n}^{\dagger}a_{m}^{\dagger}a_{k}^{\dagger}a_{n}a_{m}a_{k}|M\rangle=-
\bigg(\sum_{j=1}^M n_{j}^3\bigg)^2- \sum_{j=1}^M n_{j}^4
\sum_{k=1}^M n_{k}^2 +2\sum_{j=1}^M n_{j}^6 \ ,
\end{equation}
\begin{equation}
\langle M|\sum_{n,m,k}n^5 m
a_{n}^{\dagger}a_{m}^{\dagger}a_{k}^{\dagger}a_{n}a_{m}a_{k}|M\rangle=-(M-2)\sum_{j=1}^M
n_{j}^6 \ ,
\end{equation}
\begin{equation}
\langle M|\sum_{n,m,k}n^2 m^2 k^2
a_{n}^{\dagger}a_{m}^{\dagger}a_{k}^{\dagger}a_{n}a_{m}a_{k}|M\rangle=\bigg(\sum_{j=1}^M
n_{j}^2\bigg)^3-3 \sum_{j=1}^M n_{j}^4 \sum_{k=1}^M
n_{k}^2+2\sum_{j=1}^M n_{j}^6 \ ,
\end{equation}
\begin{equation}
\langle M|\sum_{n,m,k}n^3 m^3
a_{n}^{\dagger}a_{m}^{\dagger}a_{k}^{\dagger}a_{n}a_{m}a_{k}|M\rangle=(M-2)\bigg(\sum_{j=1}^M
n_{j}^3\bigg)^2-(M-2)\sum_{j=1}^M n_{j}^6  \ .
\end{equation}
Then the contact contribution from (\ref{3loopsix}) combined with
the exchange contribution (\ref{3loopsec}) from the second order
perturbation theory gives the  following $\tl^3/J^2$ result
\begin{eqnarray}
&&E^{(2)}_{3\ exch.+cont.} =
\frac{\tilde{\lambda}^3}{12J^2}\bigg[(62+68 \a +72
\c)\bigg(\sum_{j=1}^M n_{j}^2\bigg)^3+6(-19+14 \a -3
\b)\bigg(\sum_{j=1}^M n_{j}^3\bigg)^2\nonumber\\
&&+ \ (76 -38 \a +9 \b+144 \c) \sum_{j=1}^M n_{j}^6+( -6 -120 \a +
6 \b-216 \c)\sum_{j=1}^M n_{j}^4 \sum_{i=1}^M
n_{i}^2\nonumber\\
&&+ \ (24 \a+33)\sum_{i\ne
j}^M\frac{n_i^4n_j^4}{(n_i-n_j)^2}\bigg] \ .
 \label{3tot}
\end{eqnarray}
The final result for the $1/J^2$ correction is found  after adding
a higher-derivative term contribution discussed in the next
subsection.

\subsection{Contribution of higher-derivative terms }
\bigskip

As was already discussed in  \cite{mtt}, when  approximating
 the  discrete
spin chain coherent state action by the
 continuous   LL action one drops
certain higher-derivative corrections. These are suppressed by
$1/J$ factors in the classical
 large $J$ limit,  but they
 need to be re-instated  to correctly reproduce the spin chain
result for the $1/J^n$ contributions in the quantum LL approach.
These terms  should also be  present in the full string-theory
result,
 where they should  originate from contributions
of other  modes outside the LL  subsector (one may view the string
LL action as a result of integrating out all other superstring
world-sheet fields   while keeping $\vec n$ as a background).

The relevant higher-derivative terms in the LL action can be
obtained, e.g.,  as follows.
 The local part of all-order  dilatation
operator which contributes only to  terms quadratic in $\vec{n}$
can be represented  in the following
 symbolic way that correctly  captures the combinatorics
 of the expansion \cite{rt}
\begin{equation}\la{ddd}
D\approx \sum^J_{l=1}  \bigg( \sqrt{1+2 D_{l,l+1}}-1\bigg) \ .
\end{equation}
Here $D_{l,l+1}=I- P_{l,l+1}$ is the ``density'' of the one-loop
dilatation operator. As usual, to find the LL Lagrangian
\cite{krt,rt} one should take the coherent state expectation
value, and then the continuum limit.
 The
approximate equality in \rf{ddd}  means that  expanding the square
root
expression and taking the coherent state expectation value 
correctly reproduces the  leading order $\vec n^2$ term in the
resulting Hamiltonian. In the continuum limit
\begin{equation}
\vec n_{l+1}-\vec  n_{l}=\frac{2\pi}{J}
\partial_{1}\vec n+\frac{1}{2}\left(\frac{2\pi}{J}\right)^2
\partial_{1}^2\vec  n+\frac{1}{6} \left(\frac{2\pi}{J}\right)^3
 \partial_{1}^3
\vec n+...\  ,  \la{iio}
\end{equation}
and  $\sum^J_{l=1} \to  J \int^{2 \pi}_0 { d \s \ov 2 \pi}  $.
Ignoring total derivative terms  we have
\begin{equation}
\langle n| D_{l,l+1 }|n\rangle=\frac{\lambda}{2(4\pi)^2}   \ (\vec
n_{l+1}-\vec n_{l})^2\ \to \ - \frac{\lambda}{8J^2} \ \vec n
(\partial^2_{1} + \frac{\pi^2}{3 J^2}  \partial_{1}^{4}  +... )
\vec   n \ .
\end{equation}
Then the relevant part of the coherent state expectation value of
$D$
 may be written as  \cite{rt}
\begin{equation}\la{ded}
\langle n|D|n\rangle= J \int^{2 \pi}_0 { d \s \ov 2 \pi}  \
\frac{1}{4}  \ \vn  \
\bigg[\sqrt{1-\tilde{\lambda}\big(\partial_{1}^2+
\frac{\pi^2}{3J^2}
\partial_{1}^4 + ...\big)}-1 \bigg]\  \vn
+   O(\vec n^4) \ .
\end{equation}
This may be interpreted as quadratic part of the quantum  LL
effective action  with the exact kinetic operator  corresponding to the
exact spin-chain dispersion
relation \ci{bds}  (cf.\rf{bd})
 $\omega (n)   = \sqrt{1 + {\lambda\ov \pi^2 }  \sin^2 { p \ov 2 }} $,
 where $p= { 2  \pi \ov J} n$, or, in coordinate representation,
 $\omega ( \partial_{1} ) =
 \sqrt{1 - {\lambda\ov \pi^2 }  \sinh^2 ( {\pi  \ov  J }\partial_{1} )}
  $.
  Such exact kinetic operator is expected also to appear in the quantum
  effective action derived from full superstring theory,
  indicating that at the quantum level the superstring reveals its
  ``lattice'' or ``spin-chain'' structure.

Expanding for  large $J$ one finds
\begin{equation}
\langle n|D|n\rangle= J \int^{2 \pi}_0 { d \s \ov 2 \pi}   \bigg[\
\frac{1}{4}  \vn
\left(\sqrt{1-\tilde{\lambda}\partial_{1}^2}-1\right)\vn -
\frac{\tilde{\lambda}\pi^2}{24J^2} \ \vn
\frac{1}{\sqrt{1-\tilde{\lambda}\partial_{1}^2}}\partial_{1}^4 \vn
+... \bigg] \ .\la{ab}
\end{equation}
The leading term here is  the first term in the LL Hamiltonian in
(\ref{3loop}). To obtain the full $1/J^2$ correction one needs to
keep  the next-order term in  the above expansion \rf{ab} or add
the following term to the LL  Lagrangian
\begin{equation}
\Delta L_2 =\frac{\tilde{\lambda}\pi^2}{24J^2} \
\vn\frac{1}{\sqrt{1-\tilde{\lambda}\partial_{1}^2}}\partial_{1}^4
\vn+O({1\ov J^4}) \ .
\end{equation}
Making the field redefinition $n_a \rightarrow z_a$ in \rf{nz}, using the
momentum representation for the fluctuations of ${z_a}$ \rf{fg} (i.e. $f,g
\sim e^{in \sigma}$), and noticing that to quadratic order one can
make the replacement $\partial_{1}^2\rightarrow -n^2$, we obtain
the additional $1/J^2$ correction to the  BMN energy
\begin{equation}
\langle M|\Delta  \bar H_2
|M\rangle=-\frac{\tilde{\lambda}\pi^2}{6J^2}\sum_{i=1}^{M}\frac{n^4_i}{\sqrt{1+
\tilde{\lambda}n_{i}^2}} + O ({1 \ov J^4}) \  .
\end{equation}
This expression  matches the first term in the corresponding
energy
 (\ref{E2final}) computed  from the Bethe ansatz,
both on the gauge and the string  side. Explicitly, expanding in
powers of  $\tilde{\lambda}$ we  get
\begin{equation}
\langle M|\Delta  \bar H_2  |M\rangle=-\frac{\tl \pi^2}{6
J^2}\sum_{j=1}^{M}n_{j}^4
+\frac{\tilde{\lambda}^2\pi^2}{12J^2}\sum_{j=1}^{M}n_{j}^6-
\frac{\tilde{\lambda}^3\pi^2}{16J^2}\sum_{j=1}^{M}n_{j}^8
+O({\tilde{\lambda}^4\ov J^2})\ . \la{eess}
\end{equation}

\subsection{Final  results and discussion  }
\bigskip

We are now in a position to present the final  results for the
$1/J^2$ corrections  obtained from the quantum LL  Hamiltonian,
and to compare them  with the corresponding expressions  obtained
in Appendix B from  the Bethe ansatz  for the  gauge theory
\ci{bds}   and for the string theory \ci{afs}.

As was already mentioned, the  order $\tl/J^2$   correction was
previously found from the quantum LL approach in \ci{mtt} and was
shown to be the same as
 the one following from the Bethe ansatz.
Combining the results from (\ref{2tot}) and (\ref{eess}) we obtain
for the $\frac{\tl^2}{J^2}$   correction
\begin{equation}
  E^{(2)}_2 = \frac{\tl^2}{J^2}\left[\frac{\pi^2}{12}\sum_i^M n_i^6
-\frac{13}{2}\sum_i^M n_i^4+\frac{3}{2}\bigg(\sum_i^M
n_i^2\bigg)^2+\sum_{i\ne j}^M\frac{n_i^3n_j^3}{(n_i-n_j)^2}\right]
\ .
\end{equation}
It  is, indeed, the same as  the $\tilde{\lambda}^2/J^2$ term
found  from  both  gauge { and } string Bethe ans\"atze in
Appendix B.

Summing up  (\ref{eess}) with (\ref{3tot}) we obtain the total
expression for the $\frac{\tl^3}{J^2}$ term.  Explicitly,
 for the gauge-theory
values of the parameters $\a_{g}$, $\b_{g}$, $\c_{g}$ in \rf{opim}
we get
\begin{eqnarray}
 E^{(2)}_{3g} = && \frac{\tl^3}{J^2}\Bigg[-\frac{\pi^2}{16}\sum_i^M
n_i^8+\frac{49}{4}\sum_i^M n_i^6\nonumber\\
&&\qquad\qquad  -\ \frac{9}{4} \sum_i^M n_i^4 \sum_j^M n_j^2
-\frac{9}{2}\bigg(\sum_i^M n_i^3\bigg)^2
-\frac{1}{4}\bigg(\sum_i^M n_i^2\bigg)^3\nonumber\\
&&\qquad\qquad -\  \frac{3}{4}\sum_{i\ne
j}^M\frac{n_i^4n_j^4}{(n_i-n_j)^2}\Bigg] \ . \la{geg}
\end{eqnarray}
Remarkably, this   is   precisely the same as the result
\rf{E2final1} found  from the gauge-theory  Bethe ansatz.

For the string values $\a_{s}$, $\b_{s}$, $\c_{s}$ in \rf{oi}  we
get
\begin{eqnarray}
 E^{(2)}_{3s}= &&\frac{\tl^3}{J^2}\Bigg[-\frac{\pi^2}{16}\sum_i^M
n_i^8+\frac{49}{4}\sum_i^M n_i^6
\nonumber\\
&& \qquad\qquad -\ \frac{31}{8} \sum_i^M n_i^4 \sum_j^M
n_j^2-3\bigg(\sum_i^M n_i^3\bigg)^2+\frac{1}{8}\bigg(\sum_i^M
n_i^2\bigg)^3
\nonumber\\
&&\qquad\qquad-\ \frac{3}{4}\sum_{i\ne
j}^M\frac{n_i^4n_j^4}{(n_i-n_j)^2}\Bigg] \ . \la{ses}
\end{eqnarray}
Again, this  matches the expression \rf{E2final2} following from
the  string Bethe ansatz of \ci{afs}.

\bigskip

The conclusion that  the quantum   LL approach reproduces  the
results from gauge theory Bethe ansatz is  not totally surprising
since the $\l^3$  coefficients in the   gauge  LL  action were
essentially derived from the gauge Bethe ansatz results. We
believe our  results are  still interesting  and non-trivial
since they imply that various finite-size corrections  can be
systematically reproduced by quantizing a continuous   effective
action. Also, the matching serves as a self consistency check for
the integrability of this system.
   At $\tl^3$ order an integrable system is known that would
reproduce the Bethe equations, the Inozemtsev chain \cite{inoz}
with couplings defined as in \cite{ss}. It would be very
interesting to see if there is still matching at higher orders,
 since the spin chain that would give the Bethe
 equations in \rf{bdsBES}  and \rf{udef} is presently unknown.

The conclusion  that the quantization of the ``string'' LL action
(which is only  a  limit of the classical string action) leads to
the same result for the $1/J^2$ corrections to the BMN energies as
the string Bethe ansatz is more remarkable, since  quantum
integrability has not really been established at $\tl^3$ order for
this system, although some evidence of integrability has been
presented in \cite{beiss}. It also gives us  certain  confidence
 that  the expression \rf{ses}
is indeed the full  $\tl^3/J^2$ contribution that would follow
from the direct superstring computation (which has not yet  been
carried out
 due to its complexity, cf. \ci{Swanson}).
In addition, this  provides a strong indication that  the string
Bethe ansatz of \ci{afs} does correctly describe the quantum
$\tl^3/J^2$  string-theory
correction.  At the same time,  it also implies that  
it is only necessary to quantize the string modes that appear
explicitly in the LL action which is relevant for large $J$ in
order to reproduce the exact quantum superstring results, provided
one chooses a suitable regularization.

Our conclusions are    also consistent  with the suggestion
\ci{beiss} that the quantum string spectrum coming from the string
Bethe ansatz in \ci{afs} can be found  from a spin chain
Hamiltonian which has  order $\l^3$ coefficients for the 4-spin
terms that differ from those in the gauge-theory 3-loop dilatation
operator of \ci{bks}.\foot{We do not, however,  know how to
connect    \rf{uup} directly to the dilatation operators in
\ci{beiss}, see \ci{tse2} for a related discussion.}

\bigskip

As was first found in  \ci{callan}, the gauge and string order
$\tl^3/J$ terms in the  near-BMN energy  disagree.
 In general, for $M$
impurities one finds from the Bethe ansatz  expressions in
\ci{bds,afs} a  simple result \rf{kot} for the difference of the
string and gauge energies which can also be written as
\begin{equation}\la{koj}
E^{(1)}_{s}-E^{(1)}_{g}=
 -\frac{\tl^3}{8 J} \bigg[
\sum_i^M n_i^4 \sum_j^M n_j^2 -  \bigg(\sum_i^M n_i^3\bigg)^2
\bigg] + O(\frac{\tl^4}{J})\ .
\end{equation}
Similarly, from  the above expressions \rf{geg} and \rf{ses} we
get again a  simple expression  for the difference
\begin{equation}\la{kj}
E^{(2)}_{s}-E^{(2)}_{g}=
 -\frac{\tl^3}{8 J^2} \bigg[ 13
\sum_i^M n_i^4 \sum_j^M n_j^2 -12  \bigg(\sum_i^M n_i^3\bigg)^2+
3 \bigg(\sum_i^M n_i^2\bigg)^3 \bigg] + O(\frac{\tl^4}{J^2})\ .
\ee An explanation for the ``3-loop'' mismatches like  \rf{koj}
 was suggested in \ci{bt}.
It was observed there  that quantum superstring
 corrections to energies of {\it   ``fast'' } semiclassical strings
  contain non-analytic $\sqrt \l$
terms\foot{For ``non-fast''  semiclassical strings this, of
course, is not surprising and was known before \cite{ft1}.}
   (see also \ci{sz}).
    These should  effectively promote the coefficient
of the  quantum $\tl^3$ corrections into an interpolating function
of $\l$ which should have two different limiting  values at small
(perturbative gauge theory) and large (perturbative string theory)
values of $\l$. This also
 suggests \ci{bt}  the presence of such
interpolating functions in the $S$-matrix part
 of the string Berhe ansatz of \ci{afs}.
A similar explanation  should apply also to $1/J^2$ corrections in
\rf{kj}.

\renewcommand{\theequation}{6.\arabic{equation}}
 \setcounter{equation}{0}

\section{Non-analytic corrections}

In this  section we shall discuss  how to obtain non-analytic
terms of the type  found in \ci{bt} from the quantum LL
Hamiltonian. Here we will see that non-analytic terms are present,
but it will also be evident that other modes of the superstring
are likely to contribute to them.

The results of \ci{bt} suggest that the $1/J^2$ coefficient
function $h_2( \tl) $  in \rf{hah}  should contain  terms
  with half-integer powers of $\tl$
starting with  $\tl^{5/2}$. The  $\tl^{5/2}/J^2$  contribution
should correspond to the first subleading term in the
interpolating function $f(\l)$  in the quantum string-theory
result for the $1/J$ correction to the BMN energies. Indeed, one
can generalize \rf{kot} or \rf{koj}  to
\begin{equation}\la{ko}
E^{(1)}_{s}=E^{(1)}_{g} -\frac{\tl^3}{16 J} f(\l) \
\sum_{k,j=1}^{M} n_j^2 n_k^2 ( n_k-n_j)^2+O(\frac{\tl^4}{J})\ ,
\end{equation}
\be   f(\l)_{_{\l \gg 1}} = 1 +  {a_1 \ov \sqrt\l} + { a_2 \ov
(\sql)^2 } + ...\ , \ \  \ \ \ \ \
 f(\l)_{_{\l \to  0}}\to 0 \ . \ee
Then the  presence of the interpolating function  $f(\l)$ can
explain the ``3-loop disagreement'' found in \ci{callan}. Written
in terms of $\tl= \l/J^2$  the coefficient in \rf{ko}  is \be
\la{eex} \frac{\lambda^3}{J^7} f(\l) = {\tl^3\ov J} \bigg( 1 +
{a_1 \ov J\sqtl} +
 { a_2 \ov J^2 (\sqtl)^2 } +  ...\bigg)
\  =  {\tl^3\ov J} +  a_1 {\tl^{5/2}\ov J^2 }  + a_2 {\tl^2\ov
J^3}+
   ... \ . \ee
We  should thus expect to find the non-analytic
 $\tl^{5/2}$ term in the  string expression  for the
$M$-impurity BMN  energy at order $1/J^2$,
 and it should have a simple  coefficient proportional to
$\sum_{k,j=1}^{M} n_j^2 n_k^2( n_k-n_j)^2$. \foot{To test the
presence of other  ``non-analytic''  terms predicted by \rf{eex},
e.g., $a_2 {\tl^2\ov J^3}$,   one would  need to compute the
$1/J^3$ corrections to the BMN energies.}

The presence of the functions like  $f(\l)$ in \rf{ko} can be
related \ci{bt} to the presence of the  interpolating functions
$c_r(\l)$ in the phase part of the string Bethe ansatz
\rf{afsBES}: indeed, $c_0 (\l)$  can then be directly identified
with $f(\l)$. Assuming universality of the Bethe ansatz,   the
result of \ci{bt} about
  the 1-loop  string correction to the
 circular string state  energy then implies  that  the
first subleading coefficient in $f$ should be
  $a_1 = - { 16 \ov 3}$.

\bigskip

Our aim  below will be  to see if such non-analytic terms can be
captured in the quantum LL approach.\foot{Here we assume of course
that the starting point is the LL action with the  ``string''
coefficients in \rf{oi}:   the gauge-theory LL action viewed as an
effective action corresponding to gauge-theory spin chain should
only be treated perturbatively in $\l$.} That may seem unlikely a
priori  since in the LL action we certainly miss  some string
contributions  and so are not guaranteed to get the non-analytic
terms right; also, the issue  may be complicated by
 the presence of the  UV divergences
in the LL approach (the full string result is of course finite).
More importantly, the string LL action is  obtained by taking a
large $\J= { J\ov \sql} $  or, equivalently,  small $\tl$,
 limit of string theory,
  and its explicit all-order in $\tl$ form that generalizes  \rf{3loop}
is not known at present. However, in the case of the near-BMN
expansion it may be sufficient just to use the exact   form of the
quadratic terms in $\vn$ already included in  \rf{3loop} which are
known to  correctly reproduce the leading BMN spectrum to all
orders in $\tl$. One may then expect that using
 this  exact ``kinetic'' term
while treating other non-linear terms in \rf{3loop} perturbatively
 may be sufficient to reproduce the non-analytic
 terms in the near-BMN spectrum.

 As discussed  in \ci{bt}, the non-analytic in $\tl$  corrections in the semiclassical
 expansion  are  quantum (as opposed to ``finite-size'', cf. also \ci{pm})
 string corrections  and
 they  should  come from the large  virtual
 momenta or UV region, i.e.  they should be  present  not only
 on  an $R  \times S^1$  world-sheet
 but also on  $R^2$.
To find  non-analytic terms  in the quantum string expression one
may thus replace mode sums by momentum integrals,
 do all virtual momentum integrations and only then consider the
expansion in small $\tl$. Expanding first in $\tl$ produces (after
an appropriate regularization) only analytic corrections.

 Replacing  the sum over the  virtual quantum number $q$ by an integral in the
  exchange contribution  $ (\HH_{4}^{(1)})^2 \ov {W_{1}}$
 in \rf{secc},\rf{H42sum}\foot{All other contributions
 to the $1/J^2$ correction
 do not contain  infinite sums  and so cannot produce
 non-analytic terms.} ,
 we get
 \be \la{iin}
  Y=  \sum^M_{i\not=j}  \int^\infty_{-\infty}   d q   \
   {(\HH_{4}^{(1)})^2\ov {4W_{1}}}
  \equiv  { 1 \ov 4J^2}  \sum^M_{i\not=j} \bigg[  y(\infty) - y(-\infty) \bigg] \ . \ee
  Here we should be interested in
 the region of large $q \sim {1\ov \sqtl}  $.
 To isolate this region we may first  set $q= { x \ov \sqtl}$ and
 then  expand
 the integrand  at small $\sqtl$ for fixed  $x$.
 Performing  the indefinite integral  over $x$ we find
  a series  of terms
 \be \la{xxx}
 y(x) =   B_{-1} \tl^{-1/2}  + B_0  + B_1\tl^{1/2}   +   B_2 \tl  + B_3
  \tl^{3/2}
 + B_4  \tl^2 + B_5 \tl^{5/2} +   B_6  \tl^3 + B_7  \tl^{7/2} +  ... \ , \ee
   where the coefficients $B_k(x)$   diverge in the limit of
   large $x$.
   These singularities  represent  the UV   divergences coming from the
   bosonic fields in the $SU(2)$ LL  sector. They should be cancelled
   in the  full  superstring result against the contributions of
   other world-sheet
    fields.

    Let us recall
    that the leading $1/J$  correction  to BMN energies
  computed  from the LL  action  was also divergent and needed
  to be defined  using a particular (normal ordering and zeta-function)
  regularization in order to match the exact string/gauge theory
  results
      \ci{mtt}.
      The  complication we encounter here is that
      the above  integral  contains logarithmic
      divergencies\foot{We did not encounter
      logarithmic divergences in the computation in sect. 5.3 where we
      first expanded in $\tl$ and so it  was possible to regularize away
      all power divergences using the $\zeta$-function prescription.}
       and so it is unclear a priori how to define
      its finite part.
       Thus, starting with the quantum LL action
      we can   indeed confirm the
      presence of non-analytic contributions, but we
       are unable to compute their coefficients in an unambiguous
       way.\footnote{Similar logarithmic divergencies would appear
if one would repeat the computation of non-analytic terms in the
1-loop correction to the circular string solution  in \cite{bt} by
keeping  in the sum over characteristic frequencies only the
contributions from the 2 bosonic modes in the $SU(2)$ sector:
logarithmic divergencies are cancelled only after one includes the
contributions of all other modes.}

      Explicitly,
       using  \rf{uio},\rf{kk}, computing the integral over $x$,
    expanding
     in  large
    $x$  and symmetrising in $i,j$ we find
   \bea \la{exx}
  && B_{2k} = O( {1 \ov x})\ , \ \ \ \  k=0,1,2,3,... ,   \\
&& B_{-1} =   - \ha  - \ln ( 2 x) + O( {1 \ov x^2})  \ , \ \ \
 \ \
 B_1=  - \ha ( n_j^2 - 8 n_j n_i + n^2_i)  + O( {1 \ov x^2}) \ ,  \nonumber  \\
&& B_3 = -  n^2 _j n^2 _i\ ( 8.04 - 6.87 \ln x) +   O( {1 \ov x})\
, \nonumber \\   \ \ \ \ && B_5=  n_j^2  n_i^2\ \bigg[  6.31 n_j^2
+ 2.66 n_j n_i+ 6.31  n_i ^2
 +  \  ( 0.16  n_j^2   -    0.04 n_j n_i  + 0.16
    n_i^2) \ln x \bigg]  \nonumber\\
    &&\qquad+ \  O( {1 \ov x})\ ,
      \\
&& B_7
 = n_j^2  n_i^2\ \bigg[  2.28  n_j^4 -
        19   n_j^3 n_i+ 10 n_j^2 n_i^2 -
        19 n_j n_k^3 +
        2.28 n_i^4\nonumber \\
     &&\qquad-  (1.44 n_j^4 +
              5.39  n_j^3 n_i + 2.67  n_j^2 n_i^2 +
              5.39  n_j n_i^3 + 1.44  n_i^4)\ \ln x \bigg]
          + O( {1 \ov x})\ . \nonumber\la{gggy}
   \eea
   Here we
    ignored  all  power divergencies   but kept the
       logarithmic ones
    and did not write explicitly
     subleading $1/x^k$  terms  that do not contribute to
    \rf{iin}.

We conclude that  only the  terms with
 {\it odd }  powers of $\sqtl$
receive  non-vanishing contributions  from
 the  large $q$ region. This is
in agreement with the expectation \ci{bt} that contributions from
this  UV
 region  are responsible for the non-analytic
terms.

 We expect that  all logarithmic divergences  will get cancellled
  in the full
 string computation
 and also that the  finite parts of the coefficients $B_{2k}$
  that accompany them will get modified, so that, in particular
    the first three  coefficients $B_{-1},B_1,B_3$  will
 become zero.  If
 that happens, then indeed the   leading non-analytic term
 appearing in the string  BMN energy
 will be proportional to  $ {\tl^{5/2}\ov J^2}\sum_{i,j} B_5 (n_i,n_j) $.
These expectations are based on what happens for the  1-loop
corrections  to the energy  of the circular string solution
discussed in Appendix C.

 \bigskip

 \section{Concluding remarks}

 The computations from the string  LL action are
 relatively simple, and certainly simpler than
 the full superstring  computations
 including
 contributions from all world-sheet fields.  However, {\it a priori},
 one  cannot
  guarantee that starting from this truncated string  action
one
 will obtain
 the true superstring result for the $1/J^2$
  corrections.
  A sceptic may say
   that  both the string LL action and the quantum
   string Bethe
   ansatz  have their origin
   in the part of string action  describing   strings  already
   restricted to $R\times S^3$ and,
   in particular,
    with no fermion fields present.
   The hope, however,   is   that the effect of fermions
  and other bosonic modes  turns out to be
   relatively benign to the order considered
   and can be captured by a proper choice of the UV regularization.
This hope  was indeed realized in  several leading-order $1/J$
    computations described in
   \ci{btz,mtt}. It would
    still be an important check to carry out the full superstring
    computation of the
     $1/J^2$ correction  and  directly confirm our result   \rf{ses}.

The results presented here are also indicative of an underlying
integrability. The string LL action is derived from an $R\times
S^3$ $\sigma$-model that was
 known to be classically integrable and which effectively has
   all higher derivative
 terms included (they appear once one performs     the fast-string expansion
 and solves for the time derivatives of $\vn$, see \cite{krt,kt}).
 Here it seems that we can go a step beyond this
 and find that the string LL action is consistent with
 integrability at the first few quantum levels.   On the other hand,
  for the gauge theory we are not
 starting with a classically integrable LL action
  with all higher derivatives included, but instead
  with a proposed   two-body S-matrix with an assumed
  integrability.   Although we only have results
 to $\tl^3$ order,  it is nontrivial that one can
 construct a Lagrangian from this S-matrix that is
  local in its $\tl$ series expansion and which is
    consistent with all predictions from the Bethe ansatz.

 It is also interesting that there seem to be at least two distinct integrable LL actions.  This is in line with
 the observation in \cite{beiss} that there is
  a family of integrable dilatation operators
  at $\l^3$ order.
In fact, it might be possible to relate the two actions through
some nontrivial mapping.
 This could perhaps be accomplished using the ideas in \cite{s}.

There are several obvious generalizations.
 It would be interesting to go beyond
the $\tl^3$ order in the LL approach, especially since in
\rf{E2final} we
 already have the
all-order Bethe ansatz result. It remains to be seen if one could
effectively find the all-order form of the LL action. Another
interesting direction would be to go beyond the $SU(2)$ sector and
see if the agreement between an effective LL action and the Bethe
ansatz can be maintained.

Finally, it would be important to get a better handle on the
non-analytic terms, since it is these terms that are ultimately
responsible for the  apparent disagreements between the
perturbative gauge and
 the perturbative string expressions.

\bigskip
\bigskip

\section*{Acknowledgments }

We are  grateful to   N. Beisert, S. Frolov, R. Roiban,
S. Schafer-Nameki  and K.
Zarembo for discussions of related  issues. The work of J.A.M. was
supported in part by the Swedish Research Council. The work of
A.T. and  A.A.T.
  was supported  by the DOE grant DE-FG02-91ER40690.  A.A.T.
acknowledges  the support of
 the INTAS grant  03-51-6346 and the RS Wolfson award.

\renewcommand{\theequation}{A.\arabic{equation}}
 \setcounter{equation}{0}
\setcounter{section}{1} \setcounter{subsection}{0}

  \section*{Appendix A:
  Fixing coefficients in  3-loop gauge theory LL action:
  circular string example}

Here we  shall describe how to
 fix the coefficients \rf{opim} in the LL  action
on the gauge theory side. The coefficients $\a$ and $\b$   can be
found, e.g., by comparing the $1/J$  correction to the BMN  energy
of a generic $M$-impurity state with the Bethe ansatz result (see
sect. 4). Here we will also fix $\c$ using the simplest circular
string solution as a test  background and then check the
consistency of \rf{opim} with more general circular solutions.

 The  equal-spin $J_{1}=J_{2}$ circular string
 in the $SU(2)$ sector \cite{ft2,art} corresponds to the simplest
 non-trivial static solution of the LL action (to all orders in
 $\tl$):
\begin{equation}
\vec{n}=(\cos 2m \sigma, \ \sin 2m \sigma,\  0)\ .
\end{equation}
Here  $m$ is the  integer  winding number and   $n''_i = - 4 m^2
n_i, $ so that $n''_i  n'_i=0$. Thus its energy found from
\rf{3loop} does not depend on the value of coefficient  $\b$. The
classical energy of this solution obtained from the ``gauge" LL
Lagrangian (\ref{3loop}) is (see below)
\begin{equation}
E=J\left[1+\frac{\tilde{\lambda}}{2}m^2-\frac{\tilde{\lambda}^2}{8}m^4+
\tilde{\lambda}^3 m^6(1 + \a + \c )+O(\tilde{\lambda}^4)\right]\ .
\end{equation}
On the other hand, it was found  in \cite{ss} from the Bethe
ansatz that the three-loop correction to the anomalous dimension
of the corresponding spin chain state vanishes.
 This implies
\begin{equation} \c_g = - 1 - \a_g  \ ,  \ \ \  {\rm i.e.} \ \ \ \
\c_g=\frac{3}{4}\ ,
\end{equation}
where we used  that $\a_g= - {7\ov 4}$  as follows from  the
matching of the $1/J$ correction to the BMN energy described  in
sect. 4.

As a consistency check on the values of the $\b $ and $\c$
coefficients in \rf{opim} let us  consider the more general
circular  solution in the $SU(2)$ sector with $J_{1}\neq J_{2}$
\ci{art}. Its energy to order $\tilde{\lambda}^3$ can be computed
from the string theory \cite{art}, as well as from the gauge
theory Bethe ansatz
 \cite{m}.
Starting with the LL lagrangian (\ref{3loop}) and plugging the
leading order solution into the $\tl^3$ term in the LL
Hamiltonian  one obtains the correction to the LL energy that can
be matched onto either the string or gauge Bethe ansatz result,
thus  checking the coefficients in  \rf{oi},\rf{opim}.

Let us first recall the  details of the rational circular string
solution in the $SU(2)$ sector \cite{art}
\begin{equation}
{\rm X}_{r}=a_{r}e^{i(w_{r}\tau+m_{r}\sigma)}\ ,\ \ \ \ \ \  \quad
r=1,2
\end{equation}
where $\XX_{r}^2 =1 $ are  $S^3$ coordinates and
\begin{equation}
a_{1}^2+a_{2}^2=1\ , \quad \quad w_{r}=\sqrt{m_{r}^2+\nu^2} \ , \
\ \ \ J_r= \sqrt \l \mathcal{J}_{r} = \sqrt \l a_{r}^2w_{r}\ . \ee
$\nu$ is a parameter to be determined from the conformal gauge
constraints
\begin{equation}
\mathcal{E}^2=2 ( w_{1}\mathcal{J}_{1} +w_{2}\mathcal{J}_{2})
  -\nu^2\ , \ \ \ \ \ \ \ \quad \quad
m_{1}\mathcal{J}_{1} + m_2 \mathcal{J}_{2} =0 \ ,
\end{equation}
where the energy is $E=\sqrt{\lambda}\mathcal{E}$.
 Introducing the notation
$$m\equiv m_{1}\ , \ \ \ \ \ \
n\equiv m_{1}-m_{2}\ , \ \ \ \ \
\mathcal{J}=\mathcal{J}_{1}+\mathcal{J}_{2}\  ,
$$ one can solve one of the
constraints for $\nu$ at large $\mathcal{J}$  or small $\tl= { 1
\ov \mathcal{J}^2}$ to  obtain
\begin{equation}
\nu^2=\mathcal{J}^2 + m(m -n) - \frac{3 m(m - n)(2 m -
n)^2}{4\mathcal{J}^2} + \frac{5 m(m - n)(2 m - n)^4}{8
\mathcal{J}^4}+O({ 1 \ov \mathcal{J}^6}).
\end{equation}
Then the string energy to $\tl^3$ order is found to be
\begin{eqnarray}
&&E_s =  J \bigg[ 1 +  \frac{1}{2} \tl  m (n-m)
                          - \frac{1}{8} \tl^2  m (n-m) (n^2 - 3 m n  + 3
                          m^2)\nonumber\\
     && \ +\ \frac{1}{16} \tl^3 m (n-m) (n^4 - 7 m n^3  + 20 m^2 n^2 - 26 m^3 n + 13m^4)
                         +O(\tilde{\lambda}^4)\bigg] \ .
             \la{sti}
\end{eqnarray}
One can also compute the energy of the corresponding state
(rational one-cut solution) on  the  gauge theory side by using
the  Bethe ansatz as in \cite{m}\footnote{For this one is to plug
eq. (A.9) in \cite{m} into eq. (5.29) there.}
\begin{eqnarray}
E_g &=& J \bigg[ 1 +  \frac{1}{2} \tl  m (n-m)
                          - \frac{1}{8} \tl^2  m (n-m) (n^2 - 3 m n  + 3
                          m^2)\nonumber\\
                          &+& \frac{1}{16} \tl^3  m (n-m) (n-2m)^2(n^2 - 3 m n  + 3m^2)
                     +O(\tilde{\lambda}^4)\bigg]\ .
\la{gaui}\end{eqnarray} The difference between the two energies to
three loops has a simple form (cf.\rf{kot}, see also \ci{bt})
\begin{equation}
E_{s}-E_{g}=J \bigg[\frac{\tilde{\lambda}^3}{16}m^3
(n-m)^3+O(\tilde{\lambda}^4)\bigg]\ .
\end{equation}
Starting  now with the LL Lagrangian (\ref{3loop}) let us  find
the energy for the corresponding
 solution with $J_{1} \ne
J_{2}$  which is given to leading order by $\vn= (n_1,n_2,n_3)$
where \cite{krt}
\begin{equation}
n_{1}=2\sqrt{\frac{m}{n}\left(1-\frac{m}{n}\right)}\cos
n\sigma+O(\tilde{\lambda}), \quad \quad
n_{2}=2\sqrt{\frac{m}{n}\left(1-\frac{m}{n}\right)}\sin
n\sigma+O(\tilde{\lambda}),
\end{equation}
\begin{equation}
 n_{3}=1-\frac{2m}{n}+O(\tilde{\lambda}) \ .
\end{equation}
This solution can also be found  by expanding the full string
solution at large $\mathcal{J}$.\foot{The unit vector $\vec{n}$
can be written as $\vec{n}=(\sin 2\psi \cos 2\varphi,\sin 2\psi
\sin 2\varphi, \cos 2\psi)$. In terms of global angular
coordinates of $S^5$ with the metric $ds^2=dt^2+d\gamma^2+\cos^{2}
\gamma\ d\varphi_{3}^2+\sin^{2} \gamma\ (d\psi^2+\cos^{2}\psi\
d\varphi_{1}^2+\sin^{2}\psi\ d\varphi_{2}^2) $ we have
$\varphi=\frac{\varphi_{1}-\varphi_{2}}{2}$. Note also that the
cartesian coordinates are ${\rm X}_{1}=\cos \psi\ e^{i
\varphi_{1}}$, ${\rm X}_{2}=\sin \psi\ e^{i \varphi_{2}}.$}
Plugging this solution into the  Hamiltonian in (\ref{3loop}) we
find for its    LL  energy
\begin{eqnarray}\label{ELL}
&&E_{_{LL}}=
J\bigg[1+\frac{\tilde{\lambda}}{2}m(n-m)-\frac{\tilde{\lambda}^2}{8}m(n-m)(3m^2+n^2-3
m n)\nonumber\\
&& + \ \frac{\tilde{\lambda}^3}{16}m(n-m)\left[n^4+4 \ \a \ n^2 m
(n-m)
 + 16 \ \c \ m^2(n-m)^2  \right]+O(\tilde{\lambda}^4)\bigg]
\end{eqnarray}
One can see that for the string values $\a_{s}=-7/4$,
$\c_{s}=13/16$
 \rf{ELL} reproduces the string energy
\rf{sti}, while for the gauge values $\a_{g}=-7/4$, $\c_{g}=3/4$
\rf{ELL} reproduces the gauge theory result \rf{gaui}.

\renewcommand{\theequation}{B.\arabic{equation}}
 \setcounter{equation}{0}
\setcounter{section}{1} \setcounter{subsection}{0}
 \section*{Appendix B: $1/J^2$ corrections
 to energy of M-impurity states from the Bethe ansatz}

 In this Appendix we compute the energies of an $M$-impurity state
 up to and including $1/J^2$ corrections for the gauge  \ci{bds} and string  \ci{afs}
all-loop spin chain.

 \subsection*{B.1 Gauge theory}

  Starting with the  BDS Bethe equations \cite{bds}
 \begin{equation}\label{bdsBES}
e^{ip_iJ}=\prod_{j\ne i}^M\frac{u_i-u_j+i}{u_i-u_j-i}\,,
\end{equation}
where \be\label{udef} u=\frac12\cot\frac{p}{2}\sqrt{1+ {\l \ov
\pi^2}  \sin^2\frac{p}{2}} \,, \ \ \ \ \ \ \ \ee
 we can write (\ref{bdsBES}) up to order $1/J^2$ accuracy as
\begin{equation}\label{BES2}
e^{ip_i J}=\exp\left[\sum_{j\ne i}^M\frac{2i}{u_i-u_j}\right]\,.
\end{equation}
If we now set
\begin{equation}\label{peq}
p_i=\frac{2\pi n_i}{J}+\Delta_i\,,
\end{equation}
then  $u_i$ can be approximated as
\begin{equation}
u_i\approx  w_i\left(\frac{J}{2\pi n_i}\right)-\frac{\Delta_i}{
w_i}\left(\frac{J}{2\pi n_i}\right)^2\  , \ \ \ \ \ \ w_i \equiv
\sqrt{1+\tl n_i^2} \  ,
\end{equation}
and so $\Delta_i$ satisfies to $1/J^2$ accuracy
\begin{eqnarray}\label{Deltaeq}
\Delta_i =\frac{4\pi}{J^2}\sum_{j\ne
i}\left[\frac{n_in_j}{\th_{ji}}+\frac{J}{2\pi}\frac{n_j^2\Delta_i
w_j- n_i^2\Delta_j w_i}{ w_i w_j(\th_{ji})^2}\right], \ \ \ \ \ \
\ \ \  \th_{ji} \equiv  n_j w_i-n_i w_j \ .
\end{eqnarray}
We will assume that all $n_i$ are different. Up to the desired
order, we can write
\mbox{$\Delta_i=\Delta_i^{(1)}+\Delta_i^{(2)}$,} where
\begin{equation}
\Delta_i^{(1)}=\frac{4\pi}{J^2}\sum_{j\ne
i}^M\frac{n_in_j}{\th_{ji}}\,,
\end{equation}
and
\begin{eqnarray}
\Delta_i^{(2)}&=&
\frac{8\pi}{J^3}\Bigg[\sum_{j\ne i}^M\frac{n_in_j(n_i^2 w_i+n_j^2 w_j)}{ w_i w_j(\th_{ji})^3}\nonumber\\
&+&\sum_{k\ne j\ne i\atop k\ne i}^M\frac{n_in_jn_k(n_kn_j
w_j^2-n_kn_i w_i^2-n_j^2 w_k w_j+n_i^2 w_k w_j)}{ w_i
w_j\th_{ki}\th_{kj}(\th_{ji})^2}\Bigg]\  .
\end{eqnarray}
Now the energy is given by
\begin{eqnarray}
E&=&\sum_i^M\left(\sqrt{1+ {\l \ov  \pi^2}\sin^2\frac{p_i}{2}}-1\right)\nonumber\\
&\approx&\sum_i^M\left(\sqrt{1+  {\textstyle {\l \ov 4 \pi^2}}
p_i^2}-1-\frac{\lambda}{96 \pi^2}\frac{p_i^4}{\sqrt{1+
{\textstyle{\l \ov 4 \pi^2}} p_i^2}}\right)\,.  \la{bd}
\end{eqnarray}
Hence as an expansion in $1/J$ we find
\begin{equation}
E^{(0)}=\sum_i^M ( w_i-1)\,,
\end{equation}
\begin{equation}
E^{(1)}=\frac{J\tl}{2\pi}\sum_i ^M\frac{\Delta^{(1)}_i n_i}{
w_i}=-\frac{\tl}{J}\sum_{j\ne i}^M\frac{n_in_j}{ w_i w_j}\,,
\end{equation}
 and
\begin{eqnarray}\label{E2eq}
E^{(2)}&=&-\frac{(2\pi)^2\tl}{24J^2}\sum_i^M \frac{n_i^4}{ w_i}
+\frac{J\tl}{2\pi}\sum_i^M \frac{\Delta^{(2)}_i n_i}{ w_i}+\frac{J^2\tl}{8\pi^2}\sum_i^M \frac{(\Delta^{(1)}_i)^2}{ w_i^3}\nonumber\\
&=&-\frac{\pi^2\tl}{6J^2}\sum_i^M \frac{n_i^4}{ w_i}+\frac{4\tl}{J^2}\Bigg[\sum_{j\ne i}^M\frac{n_in_j(n_i^2 w_i+n_j^2 w_j)}{ w_i w_j(\th_{ji})^2}\nonumber\\
&+&\sum_{k\ne j\ne i\ne k}^M\frac{n_in_jn_k(n_kn_j w_j^2-n_kn_i
w_i^2-n_j^2 w_k w_j+n_i^2 w_k w_j)}{ w_i
w_j\th_{ki}\th_{kj}(\th_{ji})^2}\Bigg]
\nonumber\\
&&\qquad\qquad\qquad\qquad +\
\frac{2\tl}{J^2}\sum_{i}^M\sum_{j\ne i\atop k\ne
i}^M\frac{n_i^2n_jn_k}{ w_i^3\th_{ji}\th_{ki}}\,.
\end{eqnarray}
Symmetrizing the sums, and splitting the last term into a piece
where $k=j$ and another piece where $k\ne j$, we find
\begin{eqnarray}\label{E2final}
E^{(2)}&=&-\frac{\pi^2\tl}{6J^2}\sum_i^M \frac{n_i^4}{ w_i}
-\frac{\tl}{J^2}\sum_{j\ne i}^M\frac{n_in_j(2n_i^2 w_i^2 w_j-n_in_j( w_i^3+ w_j^3)+2n_j^2 w_j^2 w_i)}{ w_i^3 w_j^3(\th_{ji})^2}\nonumber\\
&+&\frac{2\tl}{3J^2}\sum_{k\ne j\ne i\ne k}^M\frac{n_in_jn_k}{(
w_i w_j w_k)^2\th_{ki}\th_{ij}\th_{jk}}\Bigg\{(n_i^2-n_j^2) w_i w_j \\
&&\qquad+\ (n_j^2-n_k^2) w_j w_k+(n_k^2-n_i^2) w_k w_i\nonumber\\
&&\qquad+ \  \frac{n_in_j( w_i^3- w_j^3) w_k^4+n_jn_k( w_j^3-
w_k^3) w_i^4+n_kn_i( w_k^3- w_i^3) w_i^4}{ w_i w_j
w_k}\Bigg\}\,.\nonumber
\end{eqnarray}
Note that the last two terms in \rf{E2final} are zero in the
one-loop limit, where all $ w_i=1$.

Expanding the $1/J^2$ term in \rf{E2final} to order $\tl^3$, we
find using the momentum constraint $\sum_i^M n_i=0$
\begin{eqnarray}\label{E2final1}
E^{(2)}_g&=&\frac{\tl}{J^2}\left[-\frac{\pi^2}{6}\sum_i^M n_i^4
+2\sum_i^M n_i^2 -\sum_{j\ne i}^M\frac{2n_i^2n_j^2}{ (n_i-n_j)^2}\right]\nonumber\\
&+&\frac{\tl^2}{J^2}\left[\frac{\pi^2}{12}\sum_i^M n_i^6
-\frac{13}{2}\sum_i^M n_i^4+\frac{3}{2}\left(\sum_i^M n_i^2\right)^2+\sum_{i\ne j}^M\frac{n_i^3n_j^3}{(n_i-n_j)^2}\right]\nonumber\\
&+&\frac{\tl^3}{J^2}\Bigg[-\frac{\pi^2}{16}\sum_i^M
n_i^8+\frac{49}{4}\sum_i^M n_i^6-\frac{9}{4}
\sum_i^M n_i^4  \sum_j^M n_j^2-\frac{9}{2}\left(\sum_i^M n_i^3\right)^2\nonumber\\
&&\qquad\qquad\qquad-\frac{1}{4}\left(\sum_i^M n_i^2\right)^3
-\frac{3}{4}\sum_{i\ne j}^M\frac{n_i^4n_j^4}{(n_i-n_j)^2}\Bigg]
+{\rm O}(\frac{\tl^4}{J^2})\,.
\end{eqnarray}

\bigskip

\subsection*{B.2 String theory}

Let us now consider the proposed string Bethe ansatz  in
\cite{afs}.  In this case the Bethe equations in \rf{bdsBES} are
modified to
 \begin{equation}\label{afsBES}
e^{ip_iJ}=\prod_{j\ne
i}^M\frac{u_i-u_j+i}{u_i-u_j-i}\  e^{ -2i\sum_{r=0}^\infty c_r (  \l)
\left(\frac{\l}{16 \pi^2
}\right)^{r+2}\Big[q_{r+3}(p_i)q_{r+2}(p_j)-
q_{r+3}(p_j)q_{r+2}(p_i)\Big] } \,.
\end{equation}
Below we shall ignore the contribution of the non-trivial  \ci{bt}
interpolating functions $c_r (  \l)$ setting them equal to 1.
Here $q_r(p)$ is one of the higher charges of an impurity with
momentum $p$ and is given by
\begin{equation}
q_r(p)=\frac{2\sin(\frac12(r-1)p)}{r-1}\left(\frac{\sqrt{1+
\frac{\l}{ \pi^2 } \sin^2\frac12p}-1}{ \frac{\l}{ 4\pi^2 }
\sin\frac12 p}\right)^{r-1} \,.
\end{equation}
For the accuracy desired here, we may approximate this  as
\begin{equation}
q_r(p)\approx
\left(\frac{8\pi^2}{\lambda}\right)^{r-1}\frac{1}{p^{r-2}}
{\tw}^{r-1}\,, \qquad \tw \equiv  \sqrt{1+ {\textstyle \frac{\l}{
4\pi^2 } } p^2}-1 \,.
\end{equation}
 With this approximation we can
write the extra term in \rf{afsBES} as
\begin{equation}
-2i\sum_{r=0}^\infty  \bigg[q_{r+3}(p_i)q_{r+2}(p_j)-
q_{r+3}(p_j)q_{r+2}(p_i)
\bigg]=-i\frac{\tw_i\tw_j(p_j\tw_i-p_i\tw_j)
}{p_ip_j\frac{\lambda}{4\pi^2}-\tw_i\tw_j}\,,
\end{equation}
which after much manipulation can be simplified to
\begin{equation}
i(p_i-p_j)-i\frac{p_i^2\bv_j-p_j^2\bv_i}{p_i+p_j}\,,  \qquad \
\bv_i\equiv  \sqrt{1+  {\textstyle \frac{\l}{ 4\pi^2 } } p_i^2}
\,.
\end{equation}
We thus find the equation
\begin{equation}\label{eqq}
Jp_i-2\pi n_i=\sum_{j\ne
i}^M\left(\frac{2p_ip_j}{p_j\bv_i-p_i\bv_j}+{p_i-p_j}-\frac{p_i^2\bv_j-p_j^2\bv_i}
{p_i+p_j}\right)\,.
\end{equation}
Since the extra term found in the string computation does not have
a pole at $p_i=p_j$, the residue of the double pole in
\rf{E2final} will not be affected. If one wants to keep all orders
in $\tl$, then it is convenient to replace the sum over $p_i-p_j$
in \rf{eqq} with $Mp_i$, which follows from the zero momentum
condition.   Then we find that the more natural expansion
parameter is $1/J'$ where $J'=J-M$. However, to compute terms in
the Taylor series in $\tl$, which is what we will do here, it is
easier to leave \rf{eqq} as it is. If we now substitute \rf{peq}
into \rf{eqq} we find the equation
\begin{eqnarray}\label{Delta'eq}
\Delta_i &=&\frac{4\pi}{\hJ^2}\sum_{j\ne
i}\Bigg[\frac{n_in_j}{\th_{ji}}+\frac{\hJ}{2\pi}\frac{n_j^2\Delta_i
\hw_j- n_i^2\Delta_j \hw_i}{ \hw_i \hw_j(\th_{ji})^2}\nonumber\\
&&\qquad\qquad +\
\frac12\Bigg\{n_i-n_j-\frac{n_i^2\hw_j-n_j^2\hw_i}{n_i+n_j}
+ \frac{\hJ}{2\pi}\Bigg(\Delta_i-\Delta_j\\
&&\qquad\qquad\qquad-\Delta_i\hw_j\frac{n_j^2(1-n_in_j\tl)+n_i(n_i+2n_j)\hw_i\hw_j}{\hw_i\hw_j(n+m)^2}\nonumber\\
&&\qquad\qquad\qquad+\Delta_j\hw_i\frac{n_i^2(1-n_in_j\tl)+n_j
(n_j+2n_i)\hw_i\hw_j}{\hw_i\hw_j(n+m)^2}\Bigg)\Bigg\}\Bigg]\,.
\nonumber
\end{eqnarray}
where the term in the curly brackets is the extra piece coming
from the string QBA. Again writing
$\Delta_i=\Delta_i^{(1)}+\Delta_i^{(2)}$,  we find that
\begin{equation}
\Delta_i^{(1)}=\frac{4\pi}{\hJ^2}\sum_{j\ne
i}^M\left(\frac{n_in_j}{\hth_{ji}}+\frac{n_i-n_j}{2}-\frac{n_i^2\hw_j-n_j^2\hw_i}{2(n_i+n_j)}\right)\,.
\end{equation}
The expression for $\Delta_i^{(2)}$ is quite lengthy so we will
not present the result here.  In any case,
 the expressions for $\Delta^{(1)}$ and $\Delta^{(2)}$ can
 then be plugged directly into
 the first line of \rf{E2eq}.
    The
final result for the $1/J^2$ correction to the energy, up to order
$\tl^3$ is
\begin{eqnarray}\label{E2final2}
E^{(2)}_s&=&\frac{\tl}{J^2}\left[-\frac{\pi^2}{6}\sum_i^M n_i^4
+2\sum_i^M n_i^2 -\sum_{j\ne i}^M\frac{2n_i^2n_j^2}{ (n_i-n_j)^2}\right]\nonumber\\
&+&\frac{\tl^2}{J^2}\left[\frac{\pi^2}{12}\sum_i^M n_i^6
-\frac{13}{2}\sum_i^M n_i^4+\frac{3}{2}\left(\sum_i^M n_i^2\right)^2+\sum_{i\ne j}^M\frac{n_i^3n_j^3}{(n_i-n_j)^2}\right]\nonumber\\
&+&\frac{\tl^3}{J^2}\Bigg[-\frac{\pi^2}{16}\sum_i^M
n_i^8+\frac{49}{4}\sum_i^M n_i^6-\frac{31}{8}
\sum_i^M n_i^4 \sum_j^M n_j^2 -3\left(\sum_i^M n_i^3\right)^2\nonumber\\
&&\qquad\qquad\qquad\qquad+\frac{1}{8}\left(\sum_i^M
n_i^2\right)^3-\frac{3}{4}\sum_{i\ne
j}^M\frac{n_i^4n_j^4}{(n_i-n_j)^2}\Bigg] +{\rm
O}(\frac{\tl^4}{J^2})\,.
\end{eqnarray}
Note that the order $\tl$ and $\tl^2$ terms here agree with the
gauge theory result in \rf{E2final1}, as expected.  At order
$\tl^3$ there are differences, even though there is agreement for
the pole term and the $\sum n_i^6$ and $\sum n_i^8$ terms.

\renewcommand{\theequation}{C.\arabic{equation}}
 \setcounter{equation}{0}
\setcounter{section}{1} \setcounter{subsection}{0}
 \section*{Appendix C: Non-analytic terms in one-loop
  correction to energy of
 circular string solution}

Here we shall provide some details  about the structure of
non-analytic terms in one-loop correction to the energy of
$J_1=J_2$  circular solution of \ci{ft2,art} considered in
\ci{bt}. Our motivation is to further illustrate the observation
of section 6
 that  restriction to the modes  which are present in the LL sector
  while ignoring
 all other superstring modes does
 reveal the existence of the non-analytic in $\tl$ terms
 but does
  not allow to determine them    in  an unambiguous way.

The one-loop correction to the energy has the form
\ci{fpt,btz}\foot{We consider the solution  in the form of
\ci{art}  when fermionic fluctuations are periodic (see also
\ci{btz}). In fact,
 the choice of periodicity of the fermions does not
actually influence the form of the non-analytic terms.}
 \be
 E_1={1\ov 2\k}
 \sum_{n=-\infty }^{\infty} \bigg[ \S_{_{LL}} (n,\k,m)
  + \S_{other} (n,\k,m) \bigg]
  \ .\la{E1} \ee
 Here
 \bea \la{sss}
  \S_{_{LL}} = &&\big[   n^2+2\k^2-2\rk^2
 + 2\sqrt{ (\k^2-\rk^2)^2 +n^2\k^2 } \big]^{1/2} \nn\\
 &&  +\
 \big[ n^2+2\k^2 -2\rk^2
 -2\sqrt{ (\k^2-\rk^2)^2
 +n^2\k^2} \big]^{1/2}\nn\\
&& = \sqrt{  ( n  + \sqrt{n^2 - 4 \rk^2   } )^2        + 4 \k^2
}\
 \eea
 is  the contribution of the two fluctuations from the $SU(2)$  LL sector
 (bosonic  fluctuations  in $S^3$ part of $S^5$  where the
 string is rotating, with
 third -- ``fast''-- coordinate being fixed),
 while
 \bea \la{sss1}
 \S_{other} = {4}\sqrt{ n^2+\k^2 }  +
{2}\sqrt{n^2+\k^2  -2\rk^2}-{8}\sqrt{n^2 +\k^2-\rk^2} \;
 \eea
is the contribution of 4 $AdS_5$ fluctuations ($t= \k \tau$ is not
fluctuating), 2 other modes of $S^5$ outside of $S^3$, and of 8
fermionic fluctuation modes. We use the parameters \be \la{kaj}
\k^2 = \J^2 + \rk^2  \  , \ \ \ \  \  \ \ \ \  \  \ \ \  \tl= { 1
\ov \J^2 }   \  ,  \ee where $\rk$ is an integer winding number.

Our aim is to find the terms of {\it odd} powers in $1/\J$ in the
expansion of $E_1$ at large $\J$, i.e. the terms non-analytic in
$\tl$. As discussed in \ci{bt}, it is sufficient for this to
replace the sum over  $n$ in \rf{E1} by an integral: the terms
that correct the integral to the sum happen to be  analytic in
$\tl$, i.e. contain even powers of $1/\J$ in their expansion.

Then we get $ E_1 =  \td  E_1 + $analytic terms, where \be
\la{suu}
  \td  E_1 =   \td E_{_{  LL}} +  \td E_{  other} =  {1\ov 2\k}
 \int_{-\infty }^{\infty} d n  \bigg[ \S_{_{LL}} (n,\k,\rk)
  + \S_{other} (n,\k,\rk) \bigg]  \ .
  \la{E11} \ee
Since each term in the integrand is symmetric under $n \to - n$ we
can  restrict the integral to $(0,\infty)$ and  multiply by 2.
Introducing a cutoff $ \L $  at large $n$  we then get \be
 \td  E_{_{  LL}} =
 {1\ov \k}
 \int_{0 }^{\L} d n\  \S_{_{LL}} (n,\k,\rk)   \ , \ \ \ \ \ \ \  \ \
  \td E_{  other } =
 {1\ov \k}
 \int_{0 }^{\L} d n\  \S_{other} (n,\k,\rk)   \ . \ee
 The dependence  on $\L \to \infty$ will
  of course cancels  in the sum in
\rf{suu} but we would like to study the contribution
 $
\td  E_{_{  LL}}$  separately from $\td E_{  other } $.

The integral  over $n$ in  $ \td  E_{_{  LL}}  $ can be performed
explicitly by  changing   the variable $n= {y^2 + 4 \rk^2 \ov
2y}$,  $ y= n + \sqrt{n^2  - 4 \rk^2}$. To avoid  the tachyonic
instability coming from few lowest
   LL frequencies
 we may  compute $ \td E_{_{  LL}}$  by first
  formally  setting   $\rk \to  i \rk$, computing the integral
  and rotating back in the final expression.
  \foot{By formally applying the Euler-Maclaurin formula  one can
  then check that
  the terms that correct the integral to the original  sum
  are all analytic, i.e. contain integer powers of $\tl$.}
  The imaginary part of the integral will contain only
  analytic terms, while the
non-analytic terms will be real: they are not sensitive to  this
instability having their origin in the ``stable''  large $n$
region of the spectrum.

The resulting expression is (the first square bracket is the
contribution of the upper limit $n = \L$
 in the integral  and
the second-- of the $n=0 $ point; we omit all subleading $1/\L$
contributions) \bea \la{lal} \td E_{_{  LL}} = &&  {1\ov
\k}\bigg\{ \bigg[ \L^2 + \ha \k^2
+ ( \k^2 - 2 \rk^2)  [ \ln (2 \L)   + \ln 2 ]  \bigg]\nn \\
&&-\
 \bigg[ -i \rk \sqrt{ \ka^2 - \rk^2}  +
( \k^2 - 2 \rk^2) \ln  [2 ( \sqrt{ \ka^2 -  \rk^2} + i \rk)]
\bigg] \bigg\} \ . \eea Similarly, we find for the contribution of
all other modes \bea\la{mal} \td E_{other } &=& {1\ov \k} \bigg\{
\bigg[  - \L^2  -  ( \ka^2 -  2 \rk^2) [  \ln ( 2 \L) +
 {1 \ov 2} ] \bigg] \nn \\
&-& \ \bigg[ \ka^2 \ln \ka^2 +  \ha  ( \ka^2 - 2 \rk^2) \ln (\ka^2
- 2 \rk^2)   -
  2  ( \ka^2 -  \rk^2) \ln (\ka^2 -  \rk^2)
\bigg]\bigg\} \ . \eea Summing these two contributions together we
find  that the quadratic and  logarithmic divergences indeed
cancel. Using \rf{kaj} we
 may then write the result as  the sum of the two
combinations that produce, respectively,
 even  and odd terms in the $1/\J$  expansion of $E_1$
\be \td E_1 = \td E_{even} + \td E_{odd}  \  , \ee \be \td
E_{even} = { 1 \ov  \sqrt{ \J^2  + \rk^2 } }
 \bigg[  i \J \rk  +
  \ha (\J^2  - \rk^2)
   \ln { \J - i \rk \ov \J + i \rk }  \bigg] \ ,
 \ee
\be \td  E_{odd } = { 1 \ov   \sqrt{ \J^2  + \rk^2 } }
 \bigg[   \rk^2  + 2 \J^2  \ln { \J^2  \ov \J^2 +  \rk^2}
  -  \ha   (\J^2 - \rk^2)  \ln { \J^2 -\rk^2   \ov \J^2 +  \rk^2}
   \bigg] \ ,
 \ee
where $\td  E_{odd }$ is the same as eq.(11) in \ci{bt}, so that
\be  \la{kolm} \td  E_{odd }  =  - { \rk^6  \ov 3 \J^5 } + { \rk^8
\ov 3 \J^7 } - { 49\rk^{10} \ov 120 \J^9 }   + O({1\ov  \J^{11} }
)  \ . \
 \ee
 Now, if  instead we
 concentrate    just on  the LL sector contribution in
 \rf{lal}, take its real part,
 subtract (for simplicity) the  power divergence
  but keep the logarithmic one, set
 $\L=  \J \td \L$ and expand in large $\J$ for fixed $\td \L$
 we  find:
 \bea    {\rm Re } (\td E_{_{  LL}})
&&=  {1\ov \sqrt{ \J^2 + \rk^2} } \bigg[  \ha ( \J^2 + \rk^2)
 + ( \J^2 - \rk^2) \ln (2 \L)
-   \ha   ( \J^2  -  \rk^2) \ln (\J^2 + \rk^2)    \bigg] \nn \\
&& =\ \J(\ha +  \ln (2 \td\L) )
        - { \rk^2 \ov 4 \J} ( 1 +  6\ln (2\td \L) )
 + { \rk^4 \ov 16 \J^3}  ( 15 +  14\ln (2\td \L) ) \nn \\
&&  \ \  - \ { \rk^6 \ov 96 \J^5} ( 91 +  66\ln (2\td \L) )
   + { 5\rk^8 \ov 256 \J^7} (47 + 30 \ln (2\td \L) )
 +  O({1 \ov \J^9}) \ .\la{opio}
 \eea
 Comparing this to \rf{kolm} we conclude  that the LL sector
 produces
 contributions to the odd terms that
 contain also order  $\J, 1/\J, 1/\J^3$ terms  that cancel
 against  similar terms in $\td E_{other}$ in the total
 superstring expression. The coefficients  of the higher
 $1/\J^5, 1/\J^7, ...$ terms contain  cut-off dependent parts that
 again cancel in \rf{kolm}. If one formally ignores the  singular
 $  \ln (2\td \L)$ parts,   the finite parts  of the
 $1/\J^5, 1/\J^7, ...$  coefficients in \rf{opio} are still
  different from the ones in
 \rf{kolm}, implying that {\it all}  superstring modes contribute to the
 finite parts of the coefficients of the odd (non-analitic in $\tl$)
 terms.
 These
 observations  provide support  to the  remarks
  made at the end of section 6.


\begin{thebibliography}{20}

\bi{bmn} D.~Berenstein, J.~M.~Maldacena and H.~Nastase, ``Strings
in flat space and pp waves {}from N =4 super Yang Mills,'' JHEP
{\bf 0204}, 013 (2002) [hep-th/0202021].

\bibitem{gkp}
  S.~S.~Gubser, I.~R.~Klebanov and A.~M.~Polyakov,
  ``A semi-classical limit of the gauge/string correspondence,''
  Nucl.\ Phys.\ B {\bf 636}, 99 (2002)
  [hep-th/0204051].


\bibitem{ft1}
S.~Frolov and A.~A.~Tseytlin, ``Semiclassical quantization of
rotating superstring in AdS(5) x S(5),'' JHEP {\bf 0206}, 007
(2002) [hep-th/0204226].

\bibitem{bt}
  N.~Beisert and A.~A.~Tseytlin,
  ``On quantum corrections to spinning strings and Bethe equations,''
  hep-th/0509084.


\bibitem{callan}
  C.~G. Callan, H.~K.~Lee, T.~McLoughlin, J.~H.~Schwarz, I.~Swanson and X.~Wu,
  ``Quantizing string theory in AdS(5) x S5: Beyond the pp-wave,''
  Nucl.\ Phys.\ B {\bf 673}, 3 (2003)
  [hep-th/0307032].
    C.~G. Callan, T.~McLoughlin and I.~Swanson,
  ``Holography beyond the Penrose limit,''
  Nucl.\ Phys.\ B {\bf 694}, 115 (2004)
  [hep-th/0404007].



\bibitem{ss}
  D.~Serban and M.~Staudacher,
  ``Planar N = 4 gauge theory and the Inozemtsev long range spin chain,''
  JHEP {\bf 0406}, 001 (2004)
  [hep-th/0401057].


\bi{ft2} S.~Frolov and A.~A.~Tseytlin,
  ``Multi-spin string solutions in AdS(5) x S5,''
  Nucl.\ Phys.\ B {\bf 668}, 77 (2003)
  [hep-th/0304255].

  \bibitem{parn}
   A.~Parnachev and A.~V.~Ryzhov,
   ``Strings in the near plane wave background and AdS/CFT,''
   JHEP {\bf 0210}, 066 (2002)
   [hep-th/0208010].


\bibitem{ft3}
S.~Frolov and A.~A.~Tseytlin,
  ``Quantizing three-spin string solution in AdS(5) x S5,''
  JHEP {\bf 0307}, 016 (2003)
  [hep-th/0306130].

\bibitem{fpt}
  S.~A.~Frolov, I.~Y.~Park and A.~A.~Tseytlin,
  ``On one-loop correction to energy of spinning strings in S(5),''
  Phys.\ Rev.\ D {\bf 71}, 026006 (2005)
  [hep-th/0408187].

\bibitem{ptt}
  I.~Y.~Park, A.~Tirziu and A.~A.~Tseytlin,
  ``Spinning strings in AdS(5) x S5: One-loop correction to energy in  SL(2)
  sector,''
  JHEP {\bf 0503}, 013 (2005)
  [hep-th/0501203].


\bibitem{mz}
 J.~A.~Minahan and K.~Zarembo,
 ``The Bethe-ansatz for N = 4 super Yang-Mills,''
 JHEP {\bf 0303}, 013 (2003)
 [hep-th/0212208].

\bibitem{bs}
  N.~Beisert and M.~Staudacher,
  ``The N = 4 SYM integrable super spin chain,''
  Nucl.\ Phys.\ B {\bf 670}, 439 (2003)
  [hep-th/0307042].


\bibitem{bds}
  N.~Beisert, V.~Dippel and M.~Staudacher,
  ``A novel long range spin chain and planar N = 4 super Yang-Mills,''
  JHEP {\bf 0407}, 075 (2004)
  [hep-th/0405001].


\bibitem{s}
 M.~Staudacher,
``The factorized S-matrix of CFT/AdS,'' JHEP {\bf 0505}, 054
(2005) [hep-th/0412188].

\bibitem{btz}
  N.~Beisert, A.~A.~Tseytlin and K.~Zarembo,
  ``Matching quantum strings to quantum spins: One-loop vs. finite-size
  corrections,''
  Nucl.\ Phys.\ B {\bf 715}, 190 (2005)
  [hep-th/0502173].


\bibitem{mtt} J.A. Minahan, A. Tirziu and A.A. Tseytlin,
  ``1/J corrections to semiclassical AdS/CFT states from quantum
  Landau-Lifshitz model,''
  [hep-th/0509071].


\bibitem{kru}
  M.~Kruczenski,
  ``Spin chains and string theory,''
  Phys.\ Rev.\ Lett.\  {\bf 93}, 161602 (2004)
  [hep-th/0311203].

\bibitem{krt}
  M.~Kruczenski, A.~V.~Ryzhov and A.~A.~Tseytlin,
  ``Large spin limit of AdS(5) x S5 string theory and low energy  expansion
  of ferromagnetic spin chains,''
  Nucl.\ Phys.\ B {\bf 692}, 3 (2004)
  [hep-th/0403120].

\bi{kt} M.~Kruczenski and A.~A.~Tseytlin,
  ``Semiclassical relativistic strings in S5 and long coherent operators in N
  = 4 SYM theory,''
  JHEP {\bf 0409}, 038 (2004)
  [hep-th/0406189].

\bi{inoz} V. I. Inozemtsev, ``On the connection between the
one-dimensional $S=1/2$ Heisenberg chain and Haldane-Shastry
model, J.\ Stat.\ Phys. {\bf 59}, 1143 (1990).

\bi{bks}
 N.~Beisert, C.~Kristjansen and M.~Staudacher, ``The dilatation
operator of N = 4 super Yang-Mills theory,'' Nucl.\ Phys.\ B {\bf
 664}, 131 (2003) [hep-th/0303060].



\bibitem{m}
 J.~A.~Minahan,
  ``Higher loops beyond the SU(2) sector,''
  JHEP {\bf 0410}, 053 (2004)
  [hep-th/0405243].


\bibitem{afs}
  G.~Arutyunov, S.~Frolov and M.~Staudacher,
  ``Bethe ansatz for quantum strings,''
  JHEP {\bf 0410}, 016 (2004)
  [hep-th/0406256].


\bibitem{kmmz}
 V.~A.~Kazakov, A.~Marshakov, J.~A.~Minahan and K.~Zarembo,
``Classical / quantum integrability in AdS/CFT,''
 JHEP {\bf 0405}, 024 (2004)
 [hep-th/0402207].


\bi{sz} S.~Schafer-Nameki and M.~Zamaklar,
  ``Stringy sums and corrections to the quantum string Bethe ansatz,''
  hep-th/0509096.



\bi{tse2} A.~A.~Tseytlin, ``Semiclassical strings and AdS/CFT,''
 in: Proceedings of NATO Advanced Study Institute and EC Summer
 School on String Theory: {}from Gauge Interactions to Cosmology,
 Cargese, France, 7-19 Jun 2004. hep-th/0409296.


\bibitem{rt}
  A.~V.~Ryzhov and A.~A.~Tseytlin,
  ``Towards the exact dilatation operator of N = 4 super Yang-Mills theory,''
  Nucl.\ Phys.\ B {\bf 698}, 132 (2004)
  [hep-th/0404215].


\bibitem{beisert} N.~Beisert,
  ``BMN operators and superconformal symmetry,''
  Nucl.\ Phys.\ B {\bf 659}, 79 (2003)
  [hep-th/0211032].


\bi{mets}
 R.~R.~Metsaev,
  ``Type IIB Green-Schwarz superstring in plane wave Ramond-Ramond
  background,''
  Nucl.\ Phys.\ B {\bf 625}, 70 (2002)
  [hep-th/0112044].

\bibitem{swanson}
  T.~McLoughlin and I.~Swanson,
  ``N-impurity superstring spectra near the pp-wave limit,''
  Nucl.\ Phys.\ B {\bf 702}, 86 (2004)
  [hep-th/0407240].
C.~G.~ Callan, T.~McLoughlin and I.~Swanson,
  ``Higher impurity AdS/CFT correspondence in the near-BMN limit,''
  Nucl.\ Phys.\ B {\bf 700}, 271 (2004)
  [hep-th/0405153].



\bibitem{Swanson}
  I.~Swanson,
  ``On the integrability of string theory in AdS(5) x S5,''
  hep-th/0405172.
  ``Superstring holography and integrability in AdS(5) x S5,''
  hep-th/0505028.


\bibitem{beiss}
 N.~Beisert,
  ``Spin chain for quantum strings,''
  Fortsch.\ Phys.\  {\bf 53}, 852 (2005)
  [hep-th/0409054].

\bibitem{pm}
N.~Mann and J.~Polchinski,
  ``Bethe ansatz for a quantum supercoset sigma model,''
  hep-th/0508232.

\bibitem{art}
  G.~Arutyunov, J.~Russo and A.~A.~Tseytlin,
  ``Spinning strings in AdS(5) x S5: New integrable system relations,''
  Phys.\ Rev.\ D {\bf 69}, 086009 (2004)
  [hep-th/0311004].


\end{thebibliography}
\end{document}